\documentclass[aps,showpacs,pra]{revtex4}
\usepackage{dcolumn}
\usepackage{graphicx}
\usepackage{amssymb}
\usepackage{epstopdf}
\usepackage[english]{babel}
\usepackage[utf8x]{inputenc}
\usepackage{wrapfig}
\usepackage{lipsum}
\usepackage{enumitem}

\begin{document}

\title{document}
\title{Tunable spectral narrowing enabling the functionality of graphene qubit circuits at room temperature%
}
\author{S.~E.~Shafraniuk}
\affiliation{Tegri LLC, 558 Michigan Ave, Evanston, 60202, IL, USA}
\pacs{DOI: 10.1109}

\date{\today}

\begin{abstract}
Electrically controllable quantum coherence in quantum dot clusters and arrays based on graphene stripes with zigzag atomic edges (ZZ-stripes) is studied using the Dirac equation and S-matrix technique. We find that respective multiqubit circuits promise stable operation up to room temperatures when the coherence time is prolonged up by a few orders of magnitude through the intrinsic spectral narrowing owing to electron transport between flat bands in adjacent sections. Respectively, the coupling of qubits to a noisy environment is diminished, while the inelastic electron-phonon scattering is suppressed. The Stark splitting technique enables a broad range of operations such as the all--electrical tuning of the energy level positions and width, level splitting, controlling of the inter-qubit coupling, and the coherence time. At the resonant energies, the phase coherence spreads over thousands of periods. Such phenomena potentially can be utilized in quantum computing and communication applications at room temperature.
\end{abstract}

\maketitle

\section{Introduction}
\begin{figure}
\includegraphics[width=125 mm]{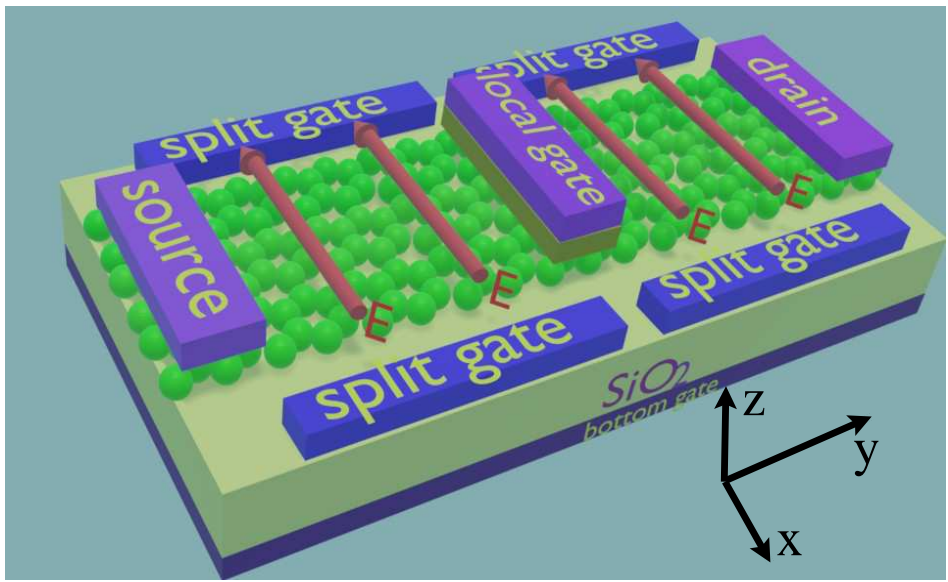} 
\caption{ {Two electrically controlled quantum dots are formed on a graphene stripe with zigzag atomic edges (ZZ-stripe). Source and drain electrodes induce electron interlevel transitions and cause interdot hoppings. The split gates create a transversal electric field ${\bf E}$ causing Stark splitting $\Delta $ of the zero-energy level whose energy diagram is also shown below in Figs.~\ref{Biased_QDot} and \ref{Fig_3apx}. The bottom gate controls the mean value of the electrochemical potential $\mu $ in the ZZ-stripe while local gates tune the height of the interdot chiral barrier $V_{\rm B}$.}}
\label{Fig_1sci}
\end{figure}

Quantum computers demonstrate impressive performance~\cite{Arute,Burnett,Kjaergaard,Krantz,Zhong,Wang,Heuck,Serh-Ivan-qubit-2002,Serhi-SISIS-qubit-2006,Serh-chapter-2008} on certain tasks that are intractable to classical computers. Quantum computers nowadays consist of dozens of relatively large qubits representing bulky, hardly scalable, and expensive devices requiring deep cooling to function properly. This motivates growing interest in the development of more powerful, compact, and cheap qubits. Although there are various suggestions for improving the quantum technology~\cite%
{Arute,Burnett,Kjaergaard,Krantz,Zhong,Wang,Heuck,Serh-Ivan-qubit-2002,Serhi-SISIS-qubit-2006,Serh-chapter-2008}, they still face many unresolved issues. One important task is creating scalable circuits comprising large arrays of tiny nanoscale qubits functioning at higher temperatures. The ultimate goal is to create an all-electrically operated large circuit, whose elements are nanoscale qubits with well-defined and tunable interlevel spacing, manageable coherence time, regulable intrinsic interqubit coupling, controllable single-photon emission, and detection.
One solution to the above problem is designing a quantum computing circuit involving quantum dot arrays (DA) based on graphene stripes~\cite{Geim-Chi-Tunn,Katsnels-Chiral-Tunnel-2012,Gold-Gordon,P-Kim,Trauzettel-Spin-qubits-GQD-2007,Shafr-Graph-Book,Fertig-1,Fertig-2,Acik-Graphene-Edges-Review-2011,Swiss-ZZ,Sh-AQT}, where coupling between the zero-dimensional localized states (LS) and the phonons occurs in a tiny phase volume. In this geometry, the thermal effects are eliminated with a controllable tuning of the dot's parameters, which is accomplished by applying appropriate gate and source-drain voltages~\cite{Sh-AQT}. As a result, the coupling of localized states to phonons is considerably reduced, and hence the thermal decoherence is largely eliminated.  Using DA where the quantum dot (QD) serves as an elementary block of the multiqubit circuit allows the all-electrical control of the interdot coupling strength, energy level number, their positions and spacing, and quantum coherence~\cite{Sh-AQT}. The major roadblock to feasible quantum computing is the insufficiently long coherence time $\tau _{\mathrm{c}}$ of QD-based qubits~\cite{Geim-Chi-Tunn,Katsnels-Chiral-Tunnel-2012,Gold-Gordon,P-Kim,Trauzettel-Spin-qubits-GQD-2007,Shafr-Graph-Book,Fertig-1,Fertig-2,Acik-Graphene-Edges-Review-2011,Swiss-ZZ,Sh-AQT}  (typically, at most $ \tau_{\rm c} \approx 10^{-10}-10^{-8}$~s even at low temperatures $T<< 300$~K). Hence, an important question is how to prolong $\tau _{\mathrm{c}}$ to achieve the flawless functionality of quantum computers. In this work, we show that $\tau _{\mathrm{c}} $ can be prolonged by several orders of magnitude in the quantum dot arrays based on graphene stripes with the zigzag shape of atomic edges (ZZ-stripe) illustrated in Fig.~\ref{Fig_1sci} instead of the using stripes with edges whose shape is the armchair~\cite{Sh-AQT}.

In this work, we examine a multiqubit system based on graphene stripe with zigzag atomic edges~\cite{Swiss-ZZ,ZZ-stripe-Carbon-2019,ZZ-stripe-topolog-insul-2011,Arabs,Serhii-Graph-THz-2019,Shafr-5}, where the electron excitation spectrum involves the edge states~\cite{Fertig-1,Fertig-2}. The quantum dots are separated from each other by chiral barriers and represent the qubits operated by applying electrical potentials to the source, drain, and local gate electrodes as illustrated in Fig.~\ref{Fig_1sci}. The height VB of chiral barriers separating sections of quantum dots, formed on the ZZ-stripe is controlled by the magnitude of the local gate voltage $V_{\rm lg}$.  Besides the interdot tunneling, the electron transport along the QD sequence involves two different types of reflection: In addition to ordinary reflection (OR) occurring due to the inter-valley backscattering process, there appears the chiral reflection (CR) caused by the intra-valley backscattering as illustrated in Fig. \ref{Fig_2apx}. The latter CR process represents an analog of Andreev reflection at the normal metal-superconductor interface. Below we will see that in QD clusters and arrays, the interdot coupling and electron energy spectrum both are controlled electrically allowing to squeeze of the energy bands and making their width $\Gamma _{n}$ exceptionally narrow. We regard this phenomenon as an \emph{intrinsic spectral narrowing}, which allows altering the qubit coherence time $\tau _{\mathrm{c}}$ on demand by prolonging it up by a few orders of magnitude, thereby improving overall functionality and performance of the multiqubit circuits.

We focus our attention on the functionality of the multiqubit circuit at elevated temperatures and discuss limitations on the coupling strength between the qubits and a noisy environment. We analyze how to extend the spatial coherence in the quantum dot array, and how to prolong the qubit's coherence time.

\section{The approach}\label{sec11}

Quantum dot arrays (DA) based on graphene stripes~\cite{Geim-Chi-Tunn,Katsnels-Chiral-Tunnel-2012,Gold-Gordon,P-Kim,Trauzettel-Spin-qubits-GQD-2007,Shafr-Graph-Book,Fertig-1,Fertig-2,Acik-Graphene-Edges-Review-2011,Swiss-ZZ,Sh-AQT} promise the remarkable potential for the operation of the DA multiqubit circuits. The aim is to considerably prolong  $\tau _{\mathrm{c}} $ by tuning the energy level width $\Gamma _{n}$ electrically. There are several factors restricting phase coherence in the quantum dots based on graphene stripes, and limiting the overall performance of the respective qubit circuits: ({\it a}) coupling to a noisy environment causes dephasing and decoherence in the multi-qubit circuit, ({\it b}) the localized electron states degrade due to the electron-impurity scattering while the electron-phonon scattering destroys the quantum coherence at elevated temperatures, ({\it c}) in the multi-dot system, the electron spectrum is rather complex, which complicates the design of multi-qubit coupling. Below we will see that the above issues can be circumvented in the electrically controllable graphene quantum dot clusters and periodic arrays based on narrow stripes with zigzag atomic edges (ZZ-stripes) as described in Appendices~\ref{secA1}, \ref{subsecA1} and \ref{subsecA2}. We implement the S-matrix technique~\cite{Datta-1995} involving the solution of the Dirac equation~(\ref{H_Dirac}) as described in Appendix~\ref{secA1}.

We use the continuity of the electron wavefunction in the QD clusters by matching it at separations of QD sections to derive the analytical form of the partial transmission $t$ and reflection $r$ coefficients as described in Appendix~\ref{subsecA2}. In analytical computing, we use the Wolfram Mathematica software, which allows finding explicit analytical forms of $t$ and $r$ [see Eqs.~(\ref{t-ZZ})-(\ref{beta-5}) in~\ref{subsecA2}]. The analytical expressions such as Eqs.~(\ref{t-ZZ})-(\ref{beta-5}) allow improving the efficiency and simplifying the numeric high precision computing in Matlab.

In Fig.~\ref{Fig_1apx} of Appendix~\ref{secA1} we sketch the flowchart of calculations involving solutions of the boundary conditions for various geometries of interest such as the single quantum dot, three- and four-dot clusters, and the periodic quantum dot array. Initially, we compute the electron dispersion law $k (\varepsilon )$ and the density of states (DOS) $N_{\rm ZZ}$ in the narrow ZZ-stripe where the edge states are formed. This allows computing of the transmission  $t (\varepsilon )$ and reflection $r (\varepsilon )$ coefficients of the $\Pi I_{G}\Pi$ block serving as an elementary part of larger clusters and arrays. Here $\Pi$ marks the space inside QD while $I_{G}$ is the voltage-controlled inter-dot barrier. Then we compute S-matrices of the $\Pi I_{G}\Pi $ block and of larger  3-dot and 4-dot "molecules" denoted as 3GM and 4GM respectively. In the next step, we compute the one-period S-matrix of an infinite periodic array comprising the one-dimensional graphene quantum dot crystal. This allows the computing of the dispersion law, the density of states, and coherence length in this system. Furthermore, we determine the dependence of level width on the interdot barrier geometry and on the type of the scattering processes causing the intrinsic spectral narrowing phenomenon facilitating the prolonged coherence time.

Solving the boundary conditions and S-matrix technique allow exact computing of the transmission and reflection coefficients using the mean-field k-p method. Furthermore, the effect of external voltage is considered exactly. Taking into account that the electron-phonon coupling constant $\lambda = 0.1 - 0.3$ is relatively small in graphene~\cite{Benedek}, as compared to metals with $\lambda \sim 0.7 - 1.4$ ~\cite{PAllen}, the scattering of electrons on phonons is considered here as weak. The external noise intensity is also considered as low.

Below we will see that the graphene qubit functionality benefits from the all-electrical control of the electron spectrum and unique transport properties of the QD clusters and arrays. In the graphene stripe with zigzag atomic edges (ZZ-stripe), one utilizes the Stark effect when the ZZ-stripe is polarized by applying the finite electric field $\mathbf{E_{\perp }}\neq 0$ in the transverse $\hat{x}$-direction. In this way, one controls the value of the Stark splitting $\Delta = e V_{\rm lg}$ by applying the electric voltage $V_{\rm lg} = \vert \mathbf{E_{\perp }}\vert W$ ($W$ is the ZZ-stripe width) to the split gate electrodes as shown in Fig.~\ref{Fig_1sci}. The origin of the edge level singularities is described in terms of the Dirac equation (\ref{H_Dirac}) for bipartite sublattices in the mean-field approximation~\cite{Shafr-Graph-Book,Ando-2005}.

\section{Intrinsic spectral narrowing of energy levels}\label{sec5}
\begin{figure}
\centering
\includegraphics[width=125 mm]{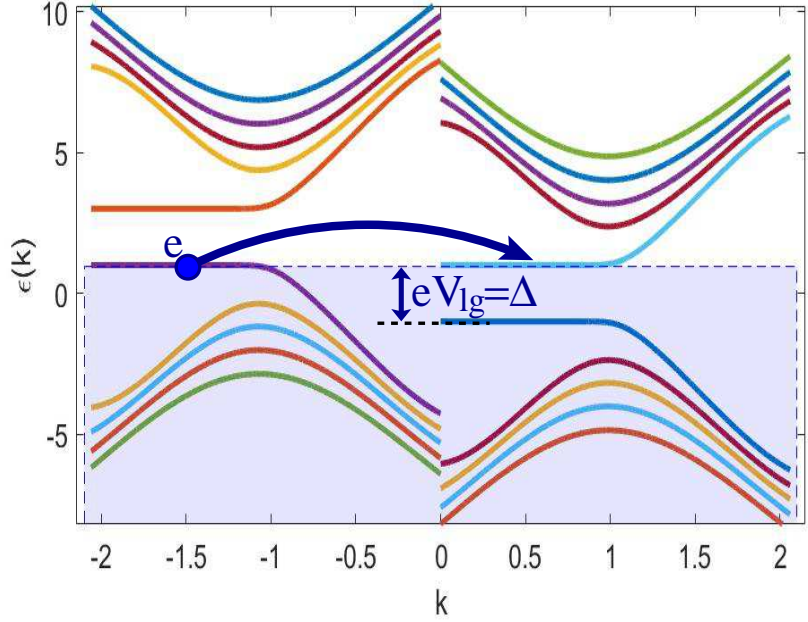} 
\caption{ {Energy diagram of electron transmission between the flat bands in neighboring sections.}}
\label{Biased_QDot}
\end{figure}

We examine the \emph{intrinsic spectral narrowing} of the energy levels representing a remarkable feature of multi-dot clusters formed on ZZ-stripes, which acts similar to conventional spectral narrowing~\cite{Asada}. We will see that in such systems, one can electrically control the energy level width $\Gamma $,  thereby tuning the coherence time $\tau _{c}$. This serves not only for intrinsic spectral narrowing when it becomes necessary during computing operations but also allows dynamic correction and optimization of the quantum computing process "on the fly".
Below we find that the level width $\Gamma $, which also determines the coherence time $\tau _{c}$, depends on the interdot coupling and on the type of elementary process. The idea is to select the desired elementary scattering process among three different types which are possible in the quantum dots based on ZZ-stripes. They involve either the scattering between two P-bands (PP-process), one P- and another F-band (PF-process) or between two F-bands (FF-process). The selection between the type of the process, that are either PP, PF, or FF is accomplished by appropriately applying electric potentials to the local source, drain, and gate electrodes, as sketched in Fig.~\ref{Fig_1sci}. Then, one readily controls the positions and spacing of energy levels localized in adjacent quantum dot sections. Remarkably, this also serves to select which type of the levels (either P or F) participates in the process of electronic inter-dot tunneling as illustrated in Fig.~\ref{Biased_QDot}. 

Electron excitation spectrum $\varepsilon (k) $ of the ZZ-stripe section subjected to transversal electric field ${\bf E_{\perp }} \neq 0$ as shown in Fig.~\ref{Fig_1sci} is presented in Appendix~\ref{secB}, Fig.~\ref{Fig_3apx}.  One can see that in addition to conventional P-bands with finite curvature, there are two flat F-bands giving rise to a large electron density of states (DOS) at $\varepsilon (k) = \pm \Delta $ (see Fig.~\ref{Fig_3apx} in Appendix) and being associated with zigzag edge states~\cite{Shafr-Graph-Book}.

In order to extract the dependence of the quantized energy level width $\Gamma $ on the type of scattering, on the interdot barrier length $L_{\mathrm{B}} $, and on the interdot barrier height $V_{\rm B}$ we consider the quantum dot clusters combining series of $\Pi I_{G}\Pi $ blocks. The calculation details are given in Appendix~\ref{subsecB1} where we describe how to compute the band structure and the electron transmission characteristics of the multi-dot quantum dot clusters and periodic arrays. During calculations, we use the electron momentum uncertainty $\delta k=0.03 \cdot K$ ( $K$ is the position of the corner of the Brillouin zone) and the respective energy uncertainty $\delta =\left( 0.01-0.03\right) \Delta $. For the sake of simplicity, we use $\delta $ as a purely phenomenological parameter (characterizing longitudinal and transverse relaxation respectively), to which many different decoherence mechanisms could, in principle, contribute. The spatial scale is introduced using that two equivalent corners of the Brillouin zone $K$ and $K^{\prime }$ are separated by $K-K^{\prime }= 4\pi /\left( 3\sqrt{3}a\right) = 9.8\times 10^{9}$~m$^{-1}$, which gives the spatial scale $2/\left(K-K^{\prime }\right) =0.2$~nm. We set Stark splitting used to solve the dispersion law as  $\Delta = 1$ (in dimensionless units), the lateral coordinate inside the ZZ-stripe $x=0.66$; the graphene ribbon width $W=14.5$, the chiral barrier length $L_{\mathrm{B}}=10.0$; the chiral barrier height $V_{\mathrm{B}} = 2.8$. 

\begin{figure}
\includegraphics[width=125 mm]{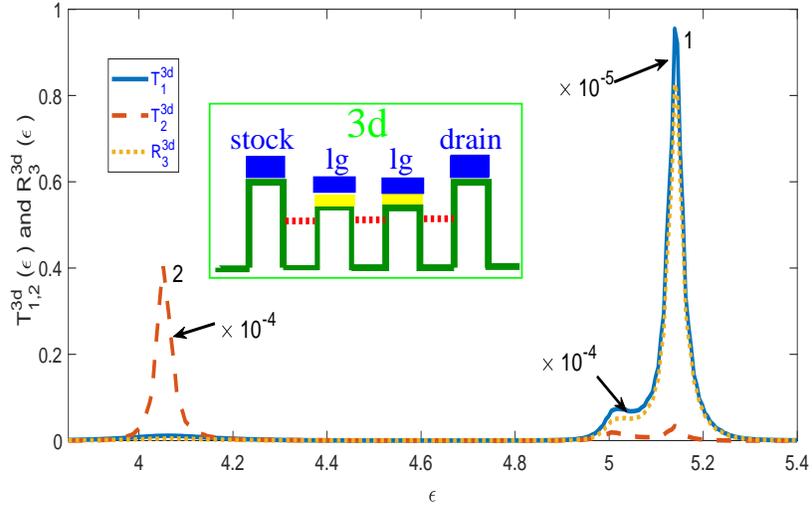} 
\caption{{The transmission and reflection probabilities through a cluster with three quantum dots where the barrier heights are $V_{\rm B} = V_{\rm lg} = 2.8$ and the electron energy uncertainty $\delta =0.03$ (in units of $\Delta $). Curves 1, 2 and 3 (yellow dots) respectively show the energy dependence of the $S_{\rm 3GM}\left( \varepsilon \right) $ components such as $T_1^{3d} = S_{\rm 3GM}^{12}$, $T_2^{3d} = S_{\rm 3GM}^{21}$ and $R_3^{3d} = S_{\rm 3GM}^{11} = S_{\rm 3GM}^{22}$ where the sharp singularities arise owing to the FF-scattering. Inset 3d shows the geometry of the three-dot cluster with stock-drain electrodes and local gates (lg) controlling the height of the interdot barriers.}}
\label{Fig_2}
\end{figure}

First, we consider an artificial "molecule" formed by three (3GM) and four (4GM) quantum dots connected in sequence and separated from each other through the chiral barriers $I_{G}$ of adjustable height. Furthermore, we compute the electron spectrum and transport properties of the infinite periodic quantum dot chain regarded here as the quantum dot crystal (GC). 
In Fig.~\ref{Fig_2} we present calculation results for the energy-dependent transmission  $T^{3d}\left( \varepsilon \right) $ and reflection $R^{3d}\left( \varepsilon \right) $ probabilities through a cluster with three quantum dots. Here we consider the chiral processes of transmission and reflection involving scattering between the flat F-bands quoted as the FF-scattering. In this Fig.~\ref{Fig_2} curves 1, 2 and 3 (yellow dots)  respectively show $T_1^{3d} = S_{\rm 3GM}^{12}$, $T_2^{3d} = S_{\rm 3GM}^{21}$ and $R_3^{3d} = S_{\rm 3GM}^{11} = S_{\rm 3GM}^{22}$ components of the respective S-matrix $S_{\rm 3GM}$, where sharp singularities arise due to the FF-scattering. Such FF-resonances are much stronger than those for PP-scattering between the conventional P-bands. 

\begin{figure}
\includegraphics[width=125 mm]{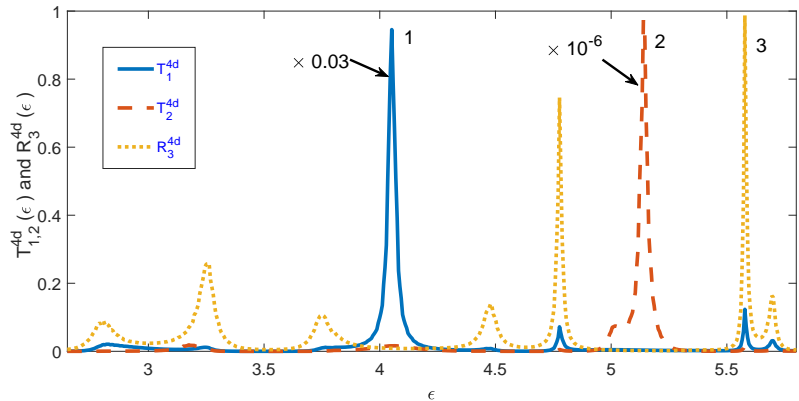} 
\caption{ {Transmission $T^{4d}_{1,2}\left( \varepsilon \right) $ (curves 1 and 2) and reflection $R^{4d}_{3}\left( \varepsilon \right) $ (curve 3) probabilities through a cluster with four quantum dots, whose parameters are the same as in Fig.~\ref{Fig_2}. Here $\varepsilon$ is the energy variable in units of Stark splitting $\Delta$. The peaks become ten times sharper and narrower as compared with the three-dot cluster. }}
\label{Fig_3}
\end{figure}

In order to further examine arising of the spectral singularities, we compute the transport and spectral characteristics for a larger 4GM "molecule", consisting of four quantum dots. The calculation results for transmission $T^{4d} $ and reflection $R^{4d} $ probabilities are presented in Fig.~\ref{Fig_3} where for the elementary block we used the same parameters as in the previous Fig.~\ref{Fig_2}. The obtained results not only confirm the effect of narrowing the spectral singularities but also suggest they become stronger for the four-dot "molecule" compared with the former three-dot counterpart: One may notice that the peak in curve 2 at $\varepsilon =5.2\Delta_n $ originating from the FF-transitions becomes ten times sharper as compared with the former one-dot and three-dot structures.

\begin{figure}
\includegraphics[width=125 mm]{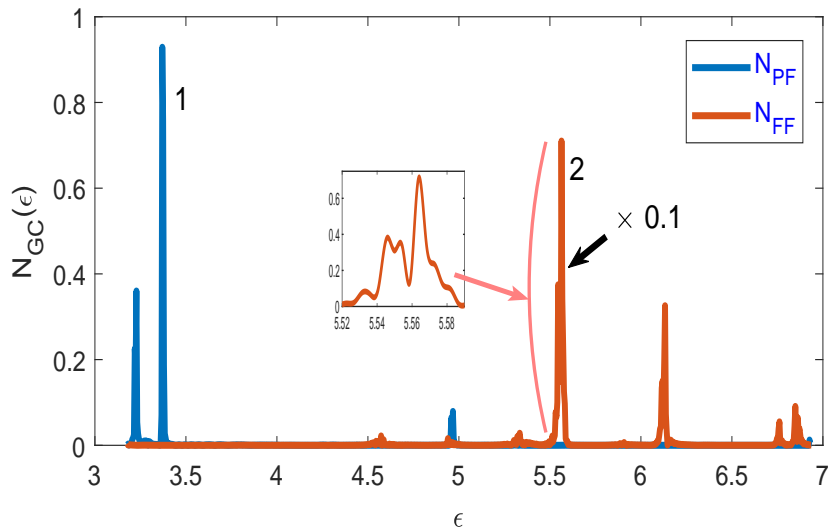} 
\caption{ {Electron density of states $N_{\rm GC}(\varepsilon )$ in the graphene quantum dot crystal. Curve 1 is for PF processes while curve 2 corresponds to FF processes. Inset shows detailed structure of the peak at $\varepsilon = 5.57 \Delta_n$. }}
\label{Fig_4}
\end{figure}

In Fig.~\ref{Fig_4} we show the electron density of states $N_{\rm GC} (\varepsilon )$ (see Eq.~(\ref{DOS}) in~\ref{subsecB2}) in the graphene quantum dot crystal (GC) comprising an infinite periodic chain of the dots formed on ZZ-stripe. The calculation details are given in Appendix~\ref{subsecB2}. One can see that $N_{\rm GC} (\varepsilon )$ represents the series of sharp peaks, whose width and magnitude depend on the type of the processes, which are either PF or FF. Remarkably, the $N_{\rm GC} (\varepsilon )$ peaks are ten times sharper and narrower for FF-processes (see curve 2 and inset) than for PF-processes (curve 1).

\begin{figure}
\includegraphics[width=125 mm]{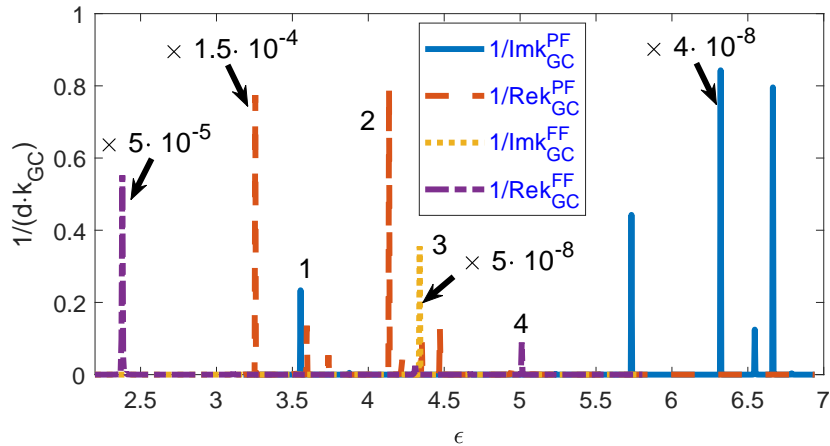} 
\caption{ {Phase coherence parameter $1/(d\cdot \kappa _{\rm GC})$ for the quantum dot crystal (GC) comprising an infinite periodic chain of the dots formed on the ZZ-stripe. Here $d$ is the GC period and $\kappa _{\rm GC}$ is the electron quasimomentum. One can see that at the resonant energies denoted by sharp peaks, the phase coherence spreads over thousands of periods.}}
\label{Fig_5sci}
\end{figure}

In Fig.~\ref{Fig_5sci} we present the numeric solution results for the phase coherence parameter $1/(d\cdot \kappa _{\rm GC}) $ in GC. The results are obtained using formulas derived in Appendix~\ref{subsecB2}. In Fig.~\ref{Fig_5sci}, $\kappa _{\rm GC}$ is the electron quasi-momentum and $d$ is the GC period [see Eq.~(\ref{kappa})]. At the resonant energies denoted by sharp peaks, the phase coherence spreads over thousand of periods, which can be utilized in quantum communication applications.

\begin{figure}
\includegraphics[width=125 mm]{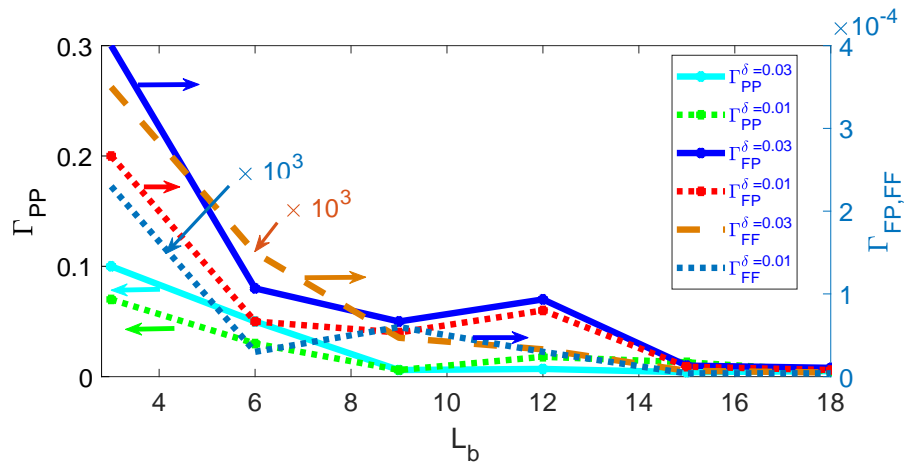} 
\caption{ {The level width $\Gamma _{\mathrm{PP,PF,FF}}$ versus the chiral barrier width $L_{\mathrm{b}}$ related to the scattering processes in the quantum dots. Cyan and green curves show $\Gamma _{\mathrm{PP}}$ for scattering between two P-bands. The electron energy uncertainties are $\delta =0.03\Delta $ and $0.01 \Delta $ respectively. Blue and red curves show $\Gamma _{\mathrm{FP}}$ for scattering between one F-band and another P-band. Finally, brown and light-blue curves show $\Gamma _{\mathrm{FF}}$ for scattering between the two flat F-bands. Here $V_{\rm lg} = \pm 0.7$, $x=0.6$.}}
\label{Fig_6}
\end{figure}

The calculation results for the three types of scattering, i.e., PP, PF, and FF are summarized in Fig.~\ref{Fig_6}. In Fig.~\ref{Fig_6} one can see that the level width $\Gamma _{\mathrm{PP}}$ for the PP-scattering is obtained as $\Gamma _{\mathrm{PP}}\simeq 10^{-3}$~eV for $\Delta = 30$~meV. The respective lifetime $\tau _{\mathrm{PP}}$ is evaluated as $\tau _{\mathrm{PP}}=\hbar /\Gamma _{\mathrm{PP}} \sim 10^{-12}$~s. For FP-scattering, the level width is reduced by the factor $\varsigma =10^{-3}$ corresponding to $\Gamma _{\mathrm{FP}}\simeq 10^{-6}$ eV , which gives $\tau _{\mathrm{FP}}\sim 10^{-9}$~s. However, for the FF-scattering $\Gamma _{\mathrm{FF}}$ is reduced again by 3-4 orders of
magnitude becoming as low as $\Gamma _{\mathrm{FF}}\simeq 10^{-9} - 10^{-9}~eV$ and hence $\tau _{\mathrm{FF}}\sim 10^{-6}-5\times 10^{-5}$~s in the latter case. Such exceptionally narrow quantized energy level peaks in ZZ-stripes occur due to the combination of three factors ({\it i}) the electron-phonon coupling is eliminated as explained in Sec.~\ref{sec12} below and in Appendix~\ref{secA01}, ({\it ii}) the interdot coupling is weak when the chiral barriers are wide and tall (i.e., for large enough $ V_{\mathrm{lg}}$ and $L_{\mathrm{B}}$) and ({\it iii}) the respective transmissions and reflections involve only the FF- scattering.  In this way one can prolong the respective $\tau _{c}$ by several orders of magnitude, e.g., from $\tau _{c}\sim 10^{-12}$~s up to $\tau _{c}\sim 5\times 10^{-5}$~s. 

The S-matrix approach (see Appendices~\ref{secA1}, \ref{subsecB2}) can be readily extended to describe the 2D quantum dot crystal by adding the periodicity condition to the second dimension. The results for the 2D case are similar to the 1D lattice, thus we omit details of the respective calculation. 

\section{Eliminating the inelastic electron-phonon scattering in the quantum dot clusters and arrays}\label{sec12}

An important characteristic of the electron transport in the graphene quantum dot structures is the electron energy relaxation time on phonons $\tau _{\mathrm{e-ph}} $  [see Eqs.~\ref{tau-BG}-\ref{lambda} in Appendix~\ref{secA01}]. The inelastic scattering time actually determines the energy dissipation in the qubit circuit and influences the coherence time $\tau _{\mathrm{c}} $ of the qubit. 

The electron-phonon coupling constant of graphene $\lambda_{\rm G} = 0.1 - 0.35$~\cite{Benedek} is far lower as compared to typical superconducting metals, where $\lambda = 0.4 - 1.3$~ \cite{PAllen}. Therefore, here we assume that the weak electron-phonon coupling in graphene is described in the linear response approximation.
General expressions for the energy-dependent $\tau (\varepsilon_k)$ and the electron-phonon decay rate $\gamma (\varepsilon_k)$ in graphene~\cite{Ando-e-ph-scatter-2009,Das-Sarma-Mobility-Graphene-2008,nanomaterials-10-00039,Phonons-Graphene-Balandin-2012,TEbook,Nika,Munoz,Sanders,Savin,Karamita} are given in Appendix~\ref{secA01}.
For the graphene quantum dots, the relevant processes of the electron-phonon scattering involve optical phonons with finite energy $ \hbar  \omega_{\rm opt}$ but zeroth momentum $q$. The phonon spectrum of graphene stripes depends on the shape of atomic edges and on the stripe width~\cite{Karamita,YWang}. In particular, for the graphene stripe with zigzag-shaped atomic edges by width $W = 1$~nm, the number of optical phonon branches per energy interval 0-50~meV is four~\cite{Karamita}, which gives the energy spacing between the optical phonon branches as $\Delta \omega_{\rm opt} \sim 12$~meV. For the wider graphene stripe by width $W = 3$~nm one gets 9 optical phonon branches spaced by 5.5~meV. The electron level spacing $\Delta_n $ varies and is controlled by applying the electric voltage to either the split gate or local gate as shown in Fig.~\ref{Fig_1sci}. Here we are interested in $\Delta_n \geq 30$~meV.

According Appendix~\ref{secA01}, one can devise a recipe for either eliminating or significantly reducing the energy dissipation due to the electron-phonon scattering. This also provides a straightforward  strategy for maintaining the long-range quantum coherence in the multi-dot structures and GC: (a) select the relevant range of electron energy and middle barrier height $V_{0}$ where just two energy levels, say $E_{1,2}$, take place, (b) use just these two energy levels, whose positions and the inter-level spacing are electrically controlled by the gate voltage, (c) furthermore, select the energy levels whose energies $E _{1,2}$ after adjustment by the phonon energy $\hbar \omega _{A_{1}^{\prime }}$ do not coincide with $E_{3,4} + \hbar \omega _{A_{1}^{\prime }}$, i.e., 
\begin{equation}
E_{3,4} + \hbar \omega _{A_{1}^{\prime }}\neq E_{1,2}.
\label{match}
\end{equation}%
The violation of the above condition~(\ref{match}) requires that
\begin{equation}
E_{3,4} + \hbar \omega _{A_{1}^{\prime }}=E_{1,2} \mbox{,}
\label{mismatch}
\end{equation}%
which means that the coherence is immediately destroyed due to the electron-phonon scattering. The latter condition~(\ref{mismatch}) can be exploited to protect a qubit against external influence or to segregate different qubits during quantum computing operations. From the above, it is clear that the electron-phonon scattering in the quantum dot clusters and arrays occurs only when the electron energy matches the condition~(\ref{mismatch}). When the condition~(\ref{mismatch}) is not satisfied, the electron-phonon interaction vanishes.

In GQD, due to the intrinsic spectral narrowing, the electron level width inside the 4-dot GQD cluster is very low, $\Gamma_{\rm FF} \simeq 1.3 \times 10^{-5}$~$\mu$eV. Since $\Gamma_{\rm FF} << \min{\{\Delta \omega_{\rm opt},\Delta _{n}\}}$, one can conform/violate the above condition~(\ref{match}) [or otherwise~(\ref{mismatch})] by merely adjusting the electron energy level positions $E_{n,m}$ (where $n$ and $m$ are the respective level indices) either by changing the interdot barrier height or by adjusting the Stark splitting magnitude as described above.
Basically, the phonon branch positions for each particular GQD configuration can also be detected experimentally by measuring the differential conductance, whose anomalies at certain values of the bias voltage would indicate when the electron bound state decay due to inelastic electron-phonon scattering occurs.

Therefore, when designing the two-qubit (or multi-qubit) gates, one should avoid the undesirable energies $E_{\rm LO/TO}=E _{3,4}-\hbar \omega _{\rm LO/TO}$ and $E_{A_{1}^{\prime }} = E _{3,4}-\hbar \omega _{A_{1}^{\prime }}$ when the dissipation becomes too large in the GM, 3GM and GC structures. One achieves this by an appropriate selection of the local gate voltage $V_{\lg }$ to ensure that the level splitting $\Delta_n $ is such that $E_{\rm LO/TO}$ and $E_{\rm{A}_{1}^{\prime }}$ don't coincide with any LS energy level. Otherwise, when there is a need to isolate the qubit, one sets the LS level to coincide with  $E_{\rm LO/TO}$ and $E_{\rm{A}_{1}^{\prime }}$. Practically, for a room-temperature functionality, one is interested in much smaller values of $\Delta \sim 30$~meV, which respectively corresponds to $W =\hbar v_{\rm F}/(\pi \Delta) \sim 20$~nm.

The total decay rate $\delta $ of the qubit's quantum state is determined by the dissipative processes of inelastic scattering and additionally by the influence of external noise,  $\delta = \Gamma_1 + \Gamma_2 + \gamma_{\rm e-ph}$. Technically, the interaction of the noise field with electrons in the quantum dot is described analogously to the electron-phonon interaction~\cite{ Zanker} (see  Appendix~\ref{secA01}). Generally, the relevant microscopic process is temperature-dependent because it represents an inelastic scattering involving changes in the electron energy and momentum. However, during the scattering in the quantum dot, the momentum and energy conservation laws impose constraints on the process probabilities, resulting in the eventual diminishing of $\delta $ due to the following. (i) Conservation of the electron momentum ${\bf p}$ requires that its change $\delta {\bf p}=0$, (ii) the energy conservation requires that the energy of a noise quantum must match the level spacing $\Delta_n $. Furthermore, in our graphene quantum dot, the electron chirality conservation introduces additional selection rules since the electron momentum and energy change during its scattering in the K-point vicinity must oblige those energy and momentum conservation rules. In this way, the temperature dependence of the dissipation processes is strongly diminished or even eliminated, protecting the graphene quantum dot qubit circuit against thermal fluctuations at elevated temperatures. The most important constraint in ZZ-qubit is the narrow width of the energy levels. This assumes that the energy dissipation does not occur unless the phonon line strictly coincides with the bound state level.

To achieve the room temperature functionality of the qubit gate, the separation between two adjacent edge state P-levels must exceed $\Delta _{n}=\hbar v_{F}/(\pi W)\approx 30$~meV provided the graphene stripe width $W=\hbar v_{F}/ (\pi \Delta_{n})<20$~nm. The qubits interact with and thus dissipate information into the noisy environment, introducing differences to the ideal result because the interaction with the environment adds a perturbation resulting in the qubit's dephasing and relaxation~\cite{Zanker}. The uncertainty due to noise arises in addition to the temperature-dependent inelastic scattering such as electron-phonon collisions resulting in the total electron energy uncertainty estimated as $\delta =  \Gamma_1 + \Gamma_2 + \gamma_{\rm e-ph} \approx \left( 0.01-0.03\right) \Delta $. The calculation details of the electron spectrum are given in Appendix~\ref{secA1}.

\section{Quantum coherence in ZZ-qubit}
Let us illustrate the preserving of quantum coherence in the graphene quantum dot qubit (ZZ-qubit) representing the open quantum system, which couples to the external environment. The ZZ-qubit is formed using the three-quantum dot cluster formed on the ZZ-stripe. We also compare the time evolution of the ZZ-qubit with another qubit, which is based on a single quantum dot. The coherence properties cannot be accurately devised by using
Schr\"{o}dinger equation
\begin{equation}
i\hbar \dot{\hat \psi} = \hat H \hat \psi \mbox{,}
\label{Schrod}
\end{equation}
where $\hat H = \hat H_0 - {\bf E}_{\lambda }(t) \cdot {\bf d}$, $ \hat H_0$ is the unperturbed Hamiltonian. Eq.~(\ref{Schrod}) describes the coupling of an optical field with electric vector ${\bf E}_{\lambda }(t)$ to the dipole moment ${\bf d}$ of the qubit that introduces time-dependent changes of the wave function $\hat \psi$. While the evolution of the state vector in a closed quantum system is deterministic, the open quantum system is stochastic in nature. The dynamics of a closed (pure) quantum system is governed by Eq.~(\ref{Schrod}), which, in principle, is solved by diagonalizing the Hamiltonian matrix $\hat H\left(t\right) $. However, $\hat H\left(t\right) $ is hard to diagonalize unless the size of the Hilbert space (dimension of the matrix $\hat H$) is small. Analytically, it is a formidable task to calculate the dynamics for systems with more than two states. If, in addition, we consider dissipation due to the unpreventable interaction with a surrounding environment, the computational complexity increases, and we have to conduct numerical calculations. The influence of an environment on the qubit causes stochastic transitions between energy levels and introduces uncertainty in the phase difference between states of the system. The state of an open quantum system is therefore described in terms of ensemble-averaged states using the density matrix formalism where the density matrix $\hat{\rho}$ describes a probability distribution of quantum states.

We use the Lindblad master equation~\cite{Breuer} governing the non-unitary time-evolution of the reduced density matrix $\hat{\rho}=\mathrm{Tr}_{\mathrm{env}}[\hat{%
\rho}_{\mathrm{tot}}]$, where env stands for environment and tot denotes the total system. In our open quantum system, incoherence is caused by the "longitudinal" and "transverse" processes characterized by the relaxation time $ T_{1} $ and dephasing time $ T_{2} $ respectively~\cite{HaugKoch}. The Lindblad equation
takes the form 
\begin{eqnarray}
\frac{d}{dt}\hat{\rho}\left( t\right)  &=&-\frac{i}{\hbar }\left[\hat H\left(
t\right) ,\hat{\rho}\left( t\right) \right] \nonumber \\
&& +\frac{\Gamma }{2}\sum_{n}\left[
2A_{n}\hat{\rho}\left( t\right) A_{n}^{\dag }-\hat{\rho}\left( t\right)
A_{n}^{\dag }A_{n}-A_{n}^{\dag }A_{n}\hat{\rho}\left( t\right) \right]
\label{LME}
\end{eqnarray}%
where $\Gamma =\hbar \left( T_{1}^{-1}+T_{2}^{-1}\right) $ is the decay rate, and $A_{n}$ are the operators through which the environment is coupled to the system. Eq.~(\ref{LME}) allows computing of the ensemble average of the system dynamics since it represents the most general trace-preserving and completely positive form describing the open system's evolution provided the decay rates $\Gamma $ are smaller than the minimum energy splitting $\Delta_n $ in the system's Hamiltonian.

We illustrate the decay using the Bloch sphere, which is a geometrical representation of the pure state space of a two-level quantum mechanical system (qubit)~\cite{Nielsen}. The Bloch sphere is a unit 2-sphere, with antipodal points corresponding to a pair of mutually orthogonal state vectors. The north and south poles of the Bloch sphere are typically chosen to correspond to the standard basis vectors $\left\vert 0\right\rangle $ and $\left\vert
1\right\rangle $. The points on the surface of the sphere correspond to the pure states of the system, whereas the interior points correspond to the mixed states. A Bloch vector $\mathbf{u}$ is a unit vector used to represent points on a Bloch sphere. When the decay rate $\Gamma \approx 0$, the Bloch vector $\mathbf{u}$ is of unit length, $|\mathbf{u}|=1$ which means that the qubit is in a pure state for all times. Obviously, because decay is always present (i.e., $\Gamma \not=0$), the qubit's density matrix, in general, is mixed and the Bloch vector has a reduced length $|\mathbf{u}|<1$. Here we use that any two-level system is dynamically equivalent to a spin--1/2 system described by the spin-vector operator $\hat{S}=\left( \hbar /2\right) \hat{\sigma}$ in terms of Pauli--matrices $\hat{\sigma}_{x}$, $\hat{\sigma}_{y}$, and $\hat{\sigma}_{z}$.

\begin{figure}
\includegraphics[width=125 mm]{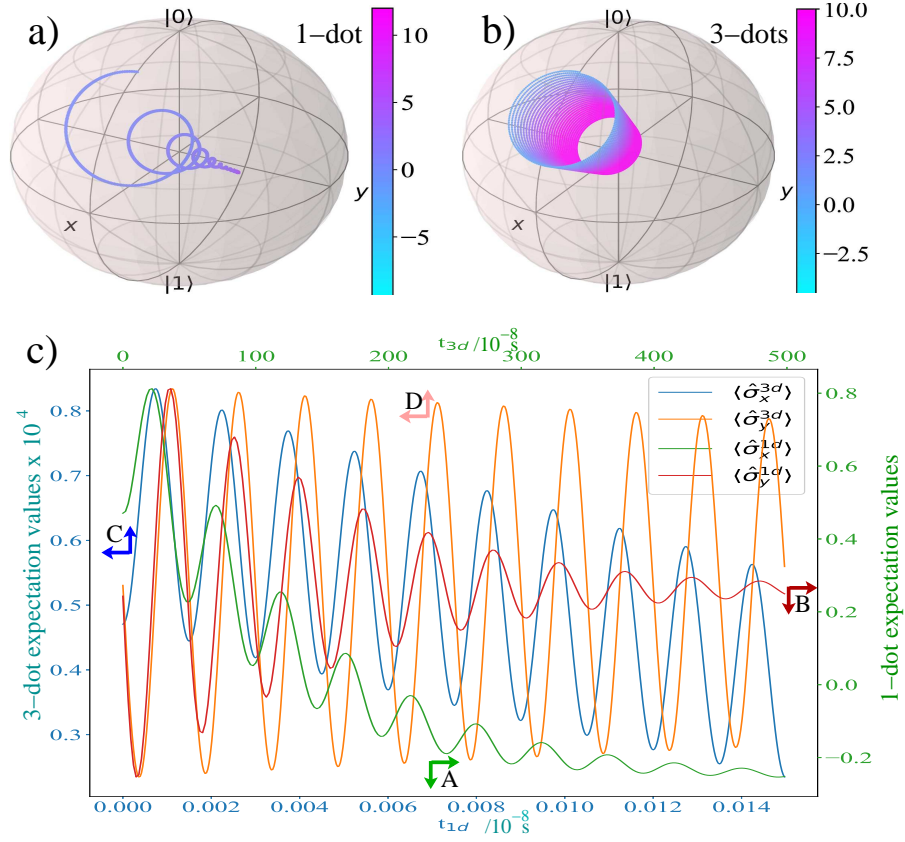} 
\caption{ {(a) The Bloch vector evolution versus time for ZZ-qubit formed of the 1-dot qubit with $\Gamma _{1}^{\mathrm{1d}}=0.0055$ and $\Gamma _{2}^{\mathrm{1d}}=0.0045$. Here and below, the energy and frequency parameters are also expressed in units of $\Delta_n $, if not stated otherwise. The qubit angle between $\hat{\sigma}_{z}$ axis and the Bloch vector is $\theta = 0.3\pi $, and $\omega _{q} = 2\pi $.  (b)~Respective plot for the 3-dot cluster with $\Gamma _{1}^{\mathrm{3d}}=5\times 10^{-7}$ and $\Gamma _{2}^{\mathrm{3d}}=4.5\times 10^{-7}$. (c) The expectation values $\left\langle \hat{\sigma}_{i}\right\rangle = \mathrm{Tr}\left[ \hat{\rho}\hat{\sigma}_{i}\right] $ ($i=x,y$) of the Bloch vector components $\hat{\sigma}_{i}$ ($i=x,y$) versus time (in units of $10^{-8}$ s) in a thermal environment characterized by $\left\langle n\right\rangle _{\mathrm{th}}=0.75$. The Rabi flopping period $T_{R}^{3d}=1.\,\allowbreak 43\times 10^{-7}~s$ in the 3-dot qubit is about four orders of magnitude longer than $T_{R}^{1d}=1.\,\allowbreak 3\times 10^{-11}~s$.}}
\label{Fig_7}
\end{figure}

Initially, when $\mathbf{E}_{\lambda }(t) \neq 0$, only the qubit is excited, but the qubit-cavity coupling results in a coherent energy transfer between the two systems, a phenomenon known as vacuum Rabi oscillations~\cite{HaugKoch}. The time evolution of the Bloch vector depends on the detuning $\delta \omega =\omega _{q}-\omega _{\lambda }$ from the resonance, where $\omega_{q}=\Delta_n /\hbar $ is the qubit angular frequency and $\omega_{\lambda }$ is the optical field frequency. For zero detuning $\delta \omega =0$ (resonance) and negligible decay $\Gamma \approx 0$, the external driving field simply causes the qubit to flop between its ground and excited states with the Rabi frequency $\Omega_{R}=\mathbf{E}_{\lambda }\cdot \mathbf{d}/\hbar $~\cite{HaugKoch}. This process is regarded as Rabi flopping. After the optical field is turned off, $\mathbf{E}_{\lambda }(t) = 0$, one observes the oscillations' decay as illustrated in Figures~\ref{Fig_7}a,b. In these Figures~\ref{Fig_7}a,b we depict the numeric solution of Eq.~(\ref{LME}) describing the time evolution of two types of qubits based on the quantum dots when the detuning is finite, $\delta \omega \not=0$. One can see the Bloch vector evolution of the 1-dot ZZ-qubit (Figure~\ref{Fig_7}a), whose coherence time is relatively short $\tau _{\mathrm{c}}^{\mathrm{1d}}\approx 50$ ps, and of the 3-dot ZZ-qubit (Figure~\ref{Fig_7}b) characterized by much longer coherence time $\tau _{\mathrm{\rm c}}^{\mathrm{3d}} = 5\times 10^{-7}$ s. They represent results obtained here as numeric solutions of Eq.~(\ref{LME}). Similar estimations for the four-dot qubit give $\tau _{\mathrm{\rm c}}^{\mathrm{3d}} = 5\times 10^{-5}$.

Next, in Figure~\ref{Fig_7}c, we compare Rabi floppings of the two qubits with the very different $\tau ^{\mathrm{1d}}_{\rm c} << \tau^{\mathrm{3d}}_{\rm c} $. The green (A) and red (B) curves are for the 1-dot qubit. For this case, we used the angular frequency $\omega _{q}^{\mathrm{1d}}=2\pi $, the qubit angle from $\hat{\sigma}_{z}$ axis was $\theta ^{\mathrm{1d}}=0.5\pi $, the qubit relaxation rate $\Gamma _{1}^{\mathrm{1d}}=0.0055$ and the dephasing rate $\Gamma _{2}^{\mathrm{1d}} = 0.0045$. The initial state is given by Eq.~(\ref{mix}) with $a=0.8$. For the 1-dot qubit, the solution of Eq.~(\ref{LME}) for $\Delta_n =30$~meV gives the qubit coherence time $\tau _{\rm c}^{\mathrm{1d}} = 50$~ps. For the other 3-dot qubit, there is a presentation issue because the fast Rabi floppings are not resolvable on the long timescale. Therefore, for illustrative reasons, we resorted to using much larger detuning (by four orders of magnitude) for the 3-dot ZZ-qubit (curves C and D) than for the 1-dot ZZ-qubit (curves A and B). This trick makes the fast Rabi flopping observable in curves C and D  in the plot on the longer time scale. Blue (C) and orange (D) curves are respectively the expectation values $\left\langle \hat{\sigma}_{x}\right\rangle $ and $\left\langle \hat{\sigma}_{y}\right\rangle $ for the 3-dot ZZ-qubit characterized by the same parameters as listed for Figure~\ref{Fig_7}b. For $\Delta_n =30$ meV, these correspond to $T_{1}^{\mathrm{3d}}=2.5\times 10^{-7}$~s and $T_{2}^{\mathrm{3d}}=3\times 10^{-7}$~s. The initial state was taken as 
\begin{equation}
\psi _{0}=(a\cdot \left\vert 2,0\right\rangle +\left( 1-a\right) \cdot
\left\vert 2,1\right\rangle )/\sqrt{a^{2}+\left( 1-a\right) ^{2}}\mbox{,}
\label{mix}
\end{equation}%
where the mixing parameter $a=0.8$. The solution of Eq.~(\ref{LME}) for $\Delta_n =30$ meV gives the 3-dot qubit coherence time as long as $\tau _{\mathrm{\rm c}}^{\mathrm{4d }}= 5\times 10^{-7}$~s. The arrows near curves A, B, C, and D point to the respective axes.

Nonetheless, the estimated value $\tau _{\mathrm{\rm c}}^{\mathrm{3d}}  = \tau _{\mathrm{FF}}\sim 10^{-7}$~s might be still too short as compared to respective values of coherence time in superconducting and trapped-ion qubits working at low temperatures. Generally speaking, the suitable value of the coherence time $\tau _{\rm c}$ depends on the operation temperature $T$ and on the "clock frequency" $f$ of the multi-qubit circuit, which can be introduced by considering the quantum devices as \textquotedblleft quantum accelerators\textquotedblright connected to a classical computer where the quantum gates are initialized by sending signals to the quantum device~\cite{Sh-AQT}. Thus, the classical process of sending a signal is clocked by a classical computer. 

We find that the ZZ-qubit functionality benefits from the all-electrical control of the electron spectrum and transport properties of the respective quantum dots. In the multi-dot qubits, one controls the level spacing $\Delta_n $ using the Stark effect by polarizing ZZ-stripe through applying a finite electric field $\mathbf{E} \neq 0$ in the transverse $\hat{x}$-direction, as shown in Fig.~\ref{Fig_1sci}. We estimate that using multi-dot qubits, flawless quantum computing at room temperatures $T \sim 300$~K can be accomplished by raising the "clock frequency" up to $f\geq 0.1$~THz.

\section{Discussion}\label{sec12}

The graphene quantum dot arrays have the potential to build large and scalable quantum computing circuits consisting of all-electrically controlled qubits having nanoscale dimensions, intrinsically coupled with each other and operating at elevated temperatures. An elementary block of the circuit represents one or several quantum dots connected in a sequence. The quantum dots are formed on a graphene stripe with atomic zigzag edges (ZZ-stripes) by width less than $5$~nm, so the edge state energy level spacing (referred here as P-bands) exceeds $\Delta _{n}\simeq \hbar v_{F}/\left( \pi W\right) =4.2\times 10^{-2}$~eV, which corresponds to the effective temperature $T^{\ast }=\Delta_{n}/k_{B}\simeq 500 $~K. If one uses Rabi frequency $f_{\mathrm{op}}\simeq 3$~THz, the respective Stark splitting of the F-bands must be $\Delta =hf_{\mathrm{op} } = 1.25\times 10^{-2}$~eV, which is achieved by applying voltage $V_{\mathrm{sg }} = 2\Delta  /\left( e\alpha _{\mathrm{sg}}\right) =0.25$~V to the split gate by efficiency $\alpha _{\mathrm{sg}}=0.1$ and corresponds to effective temperature of the QCC operation $T_{\mathrm{op}}^{\ast }\sim 300$~K. It is clear that the F-band positions are defined by Stark splitting while the P-bands depend on the stripe width $W$. Hence, when designing qubit gates, one can readily separate  F-bands and  P-bands from each other in energy just by selecting appropriate values of $V_{\rm sg}$ and $W$.

Sufficiently long coherence time $\tau_{\rm c}$ represents a key prerequisite for stable quantum computing. The phase coherence in quantum dot array (DA) is improved by reducing the coupling to a noisy environment and by turning off the temperature-dependent contribution of the inelastic collision processes electrically~\cite{Sh-AQT} (see \ref{secA01a} and \ref{secA01a}). At high temperatures $T \sim 300$~K, to insure flawless work of the qubit circuit, the worst case uncertainty of electron momentum due to coupling of qubits to the noisy environment is evaluated as $\delta k_{\mathrm{noise}} = 0.03\cdot k$. Such uncertainty arises owing to the interaction of qubits with the noise which is similar to a local interaction between electrons and phonons~\cite{Ando-e-ph-scatter-2009,Das-Sarma-Mobility-Graphene-2008,nanomaterials-10-00039,Phonons-Graphene-Balandin-2012,Sohier-thesis-2016,arXiv2004-06060v1,TEbook} and causes an uncertainty in electron energy $ \delta _{\mathrm{noise}}=\left( 0.01-0.03\right) \Delta $. Another source of decoherence at $T\approx 300$~K is the temperature-dependent inelastic scattering such as electron-phonon collisions~\cite{Sh-AQT}, which can be eliminated by applying appropriate local gate potentials. This makes the coupling to a noisy environment a dominant source of decoherence in qubit circuits.

Many of the previously developed technological solutions ensure the possibility of successful experimental realization of the proposed setup sketched in Fig.~\ref{Fig_1sci}. ({\it i})~Narrow graphene stripes with zigzag atomic edges have been obtained by various groups~\cite{Swiss-ZZ,ZZ-stripe-Carbon-2019}. ({\it ii})~The source/drain electrodes and the local top/bottom gates, which initially have been developed for carbon nanotube transistors~\cite{Rinzan,Yang,Mayle} are also used in the graphene nanodevices~\cite{Wilmart2020}. ({\it iii})~The split-gate electrodes have been used in the heat flux transistor~\cite{Mayle}.  ({\it iv})~As the local gate efficiency typically achieves $\alpha \approx 5-50\%$~\cite{Wilmart2020,Rinzan,Yang,Mayle}, one can obtain the required values of parameters (e.g., $\Delta $, $\delta $, and  $f_{\rm op}$) in relevant experiments.

There are numerous benefits to using ZZ-stripes to fabricate multi-dot devices and graphene quantum dot crystals. (a) The edge states are topologically protected~\cite{ZZ-stripe-topolog-insul-2011,Arabs} and hence they are robust against the electron scattering on lattice imperfections and phonons. (b) The electron spectrum in ZZ-stripes has two well-defined LS levels, whose spacing is readily controlled by the split gates. This greatly simplifies manipulations by the quantized states and makes the multi-qubit operations feasible. (c) The DOS peaks at the LS energies are remarkably sharp and therefore, the electron-phonon scattering is much weaker. This allows preserving a sufficient degree of quantum coherence even at elevated temperatures, opening the way to a flexible all-electrical control in the graphene quantum dot systems. The intrinsic spectral narrowing of the energy level singularities allows diminishing the level width $\Gamma $ by seven orders of magnitude. Then, by diminishing the coupling of qubits to a noisy environment and by eliminating the inelastic electron-phonon scattering one can prolong the coherence time $\tau _{\rm c} $ up to $5 \times 10^{-5}$~s even at $T\sim 300$~K. Besides, the qubit circuit functionality can be improved further by raising the gate switching frequency above $f\approx 100$~GHz, which is possible when using graphene field-effect transistors (FET)~\cite{Wilmart2020}.  This promises stable operation of respective multiqubit circuits up to room temperatures.

Traditional approaches to spectral narrowing (SN) allow the reduction of line width by about six orders of magnitude~\cite{Asada,Kitagawa,Arzi}. One solution~\cite{Asada} was based on a phase-locked loop system with a frequency-tunable oscillator integrated with a varactor diode in the slot antenna\cite{Kitagawa}. Another approach exploited the external sub-harmonic injection locking of the RTD oscillator\cite{Arzi}. In either reported cases~\cite{Asada,Kitagawa,Arzi}, the authors used an electronic circuit having a macroscopic size. However, for practical applications of the multiqubit circuits, it is highly desirable to minimize the circuit element dimensions as much as possible, which in our case becomes possible by using the intrinsic spectral narrowing based on the multi-dot clusters. This would allow fabricating of the multi-dot ZZ-qubit elements with tiny nanoscale dimensions $~20-100$~nm.

Recent experimental works~\cite{Shaikhai,Okamoto,Rinzan,Island,Dyak} on electron transport in carbon nanotubes (CNT)~\cite{Ando-2005} and in semiconducting heterostructure quantum dots exposed to THz fields suggest that such nanodevices might have a promising potential for practical applications in multi-qubit circuits. All-electrical control of the underlying circuit allows engineering with trade-offs for various qubit characteristics, such as transition frequency, anharmonicity, and sensitivity to various noise sources. In this way, highly tailored quantum programs take advantage of detailed knowledge of a given quantum device~\cite{Shaikhai,Okamoto,Rinzan,Island,Dyak}. Such benefits of GC based on ZZ-stripe provide a potential solution when evaluated coherence time $\tau _{\rm c} \approx 5 \times 10^{-5}$~s would be sufficient to maintain flawless work of the high-speed quantum computing circuits even at $T \approx 300$~K provided $f \geq 0.1$~THz.

The quantum dot clusters and crystals can host either charge or long-lived spin qubits. As a charge qubit, they function when a double-dot system (DDS) operates in the single electron regime provided one electron is shared between the two quantum dots. Otherwise, when DDS works as a singlet or triplet qubit in the two-electron regime, the system represents a spin qubit. The total capacitance, which depends on external variables such as energy detuning, temperature, and magnetic field determines the type of the system. There are two relevant components of the qubit's capacitance: i) Quantum capacitance caused by the adiabatic charge transitions and by the non-zero curvature of the energy bands and ii) the tunneling capacitance determined by the population redistribution processes, such as resonant excitation or relaxation, taking place when the rate exceeds the probing frequency.

\section{Conclusion}\label{sec13}
In summarizing, a novel concept of device geometry shown in Fig.~\ref{Fig_1sci} is proposed to allow electrically narrowing of the spectrum and enable room-temperature operation of graphene multi-qubit circuits. In this geometry, a transverse electric field is used to create F-bands, which result in sharp singularities in the transmission/reflection probabilities curve due to FF-scattering. Furthermore, the local gate voltage is used to create the chiral barrier and to optimize FF-scattering between the quantum dots, thereby reducing level width and providing a long coherence time. We have found that the suggested approach potentially can solve the decoherence issues by prolonging the coherence time and length by seven orders of magnitude, thereby opening the path to creating portable quantum computers, THz lasers, quantum detectors, and quantum communication functioning at elevated temperatures. 

\section{Acknowledgments}

The author wishes to thank Ivan Kravchenko and Dennis Drew for extremely valuable discussions.

\appendix

\section{Electron spectrum and transport in graphene quantum dot clusters and periodic arrays}\label{secA1}

\begin{figure}
\centering
\includegraphics[width=125 mm]{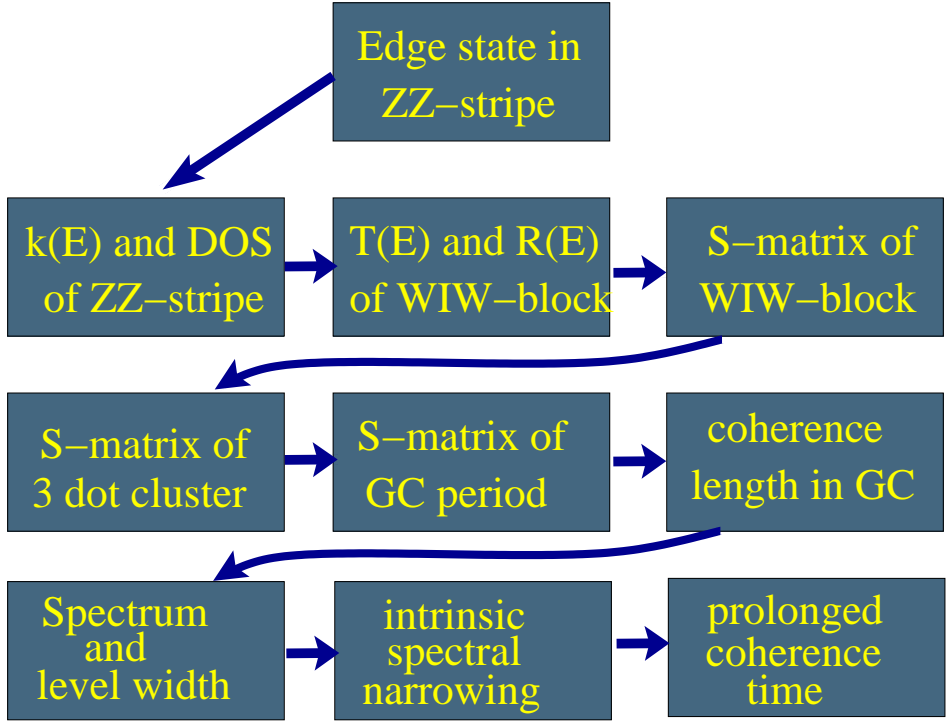} 
\caption{ {Modeling flowchart of the quantum dot clusters and arrays based on ZZ-stripes.}}
\label{Fig_1apx}
\end{figure}

The modeling step-by-step flowchart of the quantum dot clusters and arrays based on ZZ-stripes is shown in Fig.~\ref{Fig_1apx}.

Properties of quantum dots formed on ZZ-stripe are described in terms of the Dirac equation. To find the Stark splitting of zero-energy energy level, one adds a symmetry-breaking term $\propto \Delta $ into the Dirac Hamiltonian (See Ref.~\cite{Shafr-Graph-Book}, Secs.~3.3-3.5)%
\begin{eqnarray}
\mathcal{H} &=&-i\hbar v\left( \left( \hat{\sigma}_{x}\otimes \hat{1}\right)
\partial _{x}+\left( \hat{\sigma}_{y}\otimes \hat{\tau}_{z}\right) \partial
_{y}\right)  \nonumber \\
&&+V\left( x\right) \left( \hat{1}\otimes \hat{1}\right) +\Delta \left( \hat{%
\sigma}_{z}\otimes \hat{\tau}_{z}\right) ,  \label{H_Dirac}
\end{eqnarray}%
where $v=8.1\times 10^{5}$m/s$\simeq c/300$ is the massless fermion speed, $\hat{\sigma}_{i}$ and $\hat{\tau}_{k}$ are the Pauli matrices, $\otimes $ is the Kronecker product, $\{i,k\}=1\dots 3$, and Stark level splitting term~
\cite{Shafr-Graph-Book} is $\Delta =e \vert \mathbf{E}\vert W/2=e\alpha _{\mathrm{sg }}V_{\mathrm{sg}}/2$, $V_{\mathrm{sg}}$ is the split gate voltage, $\alpha _{%
\mathrm{sg}}=0.02-0.4$ is the split gate efficiency. 
When computing the electron excitation spectrum of the ZZ-stripe, we neglect the effect of decay $\delta $ on the electron energy, which is small even for the conventional single quantum dot, where we evaluate $\delta = 2~\mu$eV, which is far lower than the level spacing of interest $\Delta = 30$~meV. For ZZ-qubit, we estimate $\delta =  2 \times 10^{-10}$~eV, which is even lower. Therefore, in beginning, small corrections to the electron energy level shift due to the external noise can be disregarded, while the influence of noise on the decay is essential as explained below in calculations of \ref{secA1} and \ref{secB}.
For the stripe with \emph{zigzag} atomic edges (ZZ-stripe) the
boundary conditions are $\Psi _{A}^{\left( K\right) }\left( 0\right) =\Psi_{A}^{\left( K^{\prime }\right) }\left( 0\right) =0$ at $y=0$ and $\Psi_{A}^{\left( K\right) }\left( W\right) =\Psi _{A}^{\left( K^{\prime }\right) }\left( W\right) =0$ at $y=W$, which gives the transcendental equation~\cite{Fertig-1,Fertig-2} 
\begin{equation}
k=\frac{q_{n}}{\tan \left( q_{n}W\right) },  \label{Fert_BC_B}
\end{equation}%
where now
\begin{equation}
q_{n}=\pm \sqrt{\left( \varepsilon /(\hbar v_{\mathrm{F}})\right)
^{2}-k^{2}-\left( \Delta /(\hbar v_{\mathrm{F}})\right) ^{2}}.
\label{dispers}
\end{equation}%

\subsection{Electron spectrum and transport characteristics of a single block}\label{subsecA1}

Below,  using the piece-wise approximation, we compute the electron transport characteristics of the graphene quantum dot clusters and arrays (see scheme in Fig.~\ref{Fig_1apx}). Initially, we calculate the coefficients $t$ and $r$ of transmission and reflection for elementary chiral barrier and well. In the next steps, they are used for obtaining the S-matrices~\cite{Datta-1995} of more complex blocks and structures composed of quantum dots and representing artificial atoms (GA), their combinations quoted as "molecules" (GM), and infinite periodic quantum dot arrays regarded as artificial crystals (GC). An example of the two-dot 2GM "molecule" is shown in Fig.~\ref{Fig_1sci}.

The elementary block comprises three sections $I\Pi I_{G}$, $\Pi I_{G}\Pi $ and $I_{G}\Pi I$, where $I$ stands for the barrier of fixed height $V_{\mathrm{B}}$, $\Pi $ is the dot region and $I_{G}$ is the barrier, whose height $V_{\mathrm{B}}$ is controlled by applying the local gate voltage $V_{\rm lg}$. Respective shifts $%
V_{\rm lg}^{\left( i\right) }$ ($i=1,2,3$) of the electrochemical potential in the separate sections 1, 2 and 3 of an elementary block are characterized by a piece-wise potential
\begin{equation}
V_{\rm lg}\left( y\right) =\left\{ 
\begin{array}{c}
V_{\rm lg}^{\left( 1\right) }\mbox{ at }y<y_{\mathrm{L}} \\ 
V_{\rm lg}^{\left( 2\right) }\mbox{ at }y_{\mathrm{L}}<y<y_{\mathrm{R}} \\ 
V_{\rm lg}^{\left( 3\right) }\mbox{ at }y>y_{\mathrm{R}},%
\end{array}%
\right\vert  \label{V-y}
\end{equation}%
where $y_{\mathrm{L,R}}$ are the coordinates of the left and right edges of the chiral barrier respectively. ln the sections 1, 2, and 3, the electron envelope wavefunction $\Phi \left( x,y\right) $ is represented by solutions of Dirac equation. 

The dependence of $\Phi \left( x,y\right) $ on the transverse
coordinate $x$ in each section is approximated using the solution of Refs.~\cite{Fertig-1,Fertig-2}. Within the section $\Pi $ of a narrow graphene stripe of width $W$, when the longitudinal momentum $k>k_{\rm c}=1/W$ and the transversal electron momentum $q=i\sqrt{k^{2}-\varepsilon _{p}^{2}/v^{2}}$ becomes purely imaginary $q=\chi $, the electron excitation spectrum is given by~\cite%
{Fertig-1,Fertig-2}%
\begin{equation}
\frac{k-\chi }{k+\chi }=e^{-2W\chi }.  \label{Fert_BC}
\end{equation}%
Instead, for real $q=\sqrt{\varepsilon _{p}^{2}-k^{2}}$ one obtains%
\begin{equation}
k=\frac{q_{n}}{\tan \left( q_{n}W\right) },  \label{Fert_BC_C}
\end{equation}%
which gives quantized states in the transversal direction with the energies $\varepsilon _{k}^{\left( n\right) }=vp_{n}=v\sqrt{q_{n}^{2}+k^{2}}$, where $\varepsilon _{p} $ is the electron excitation energy, $v\simeq 10^{6}$~m/s is the Fermi velocity in graphene, and $p_{n}$ is the absolute value of the
electron momentum. The longitudinal electron momentum $k_{i}$ ($i=$ L,M,R) is obtained as a solution of the transcendental equation (\ref{Fert_BC_C}), where now $q_n = \sqrt{\left( \varepsilon _{k}^{\left( n\right) }-V_{\rm G}\right)^{2}/v^{2}-k_{i}^{2}-\Delta _{i}^{2}/v^{2}}$ where $V_{\rm G}$ is the bottom gate potential that is applied as shown in Fig.~\ref{Fig_1sci}.

\begin{figure}
\centering
\includegraphics[width=125 mm]{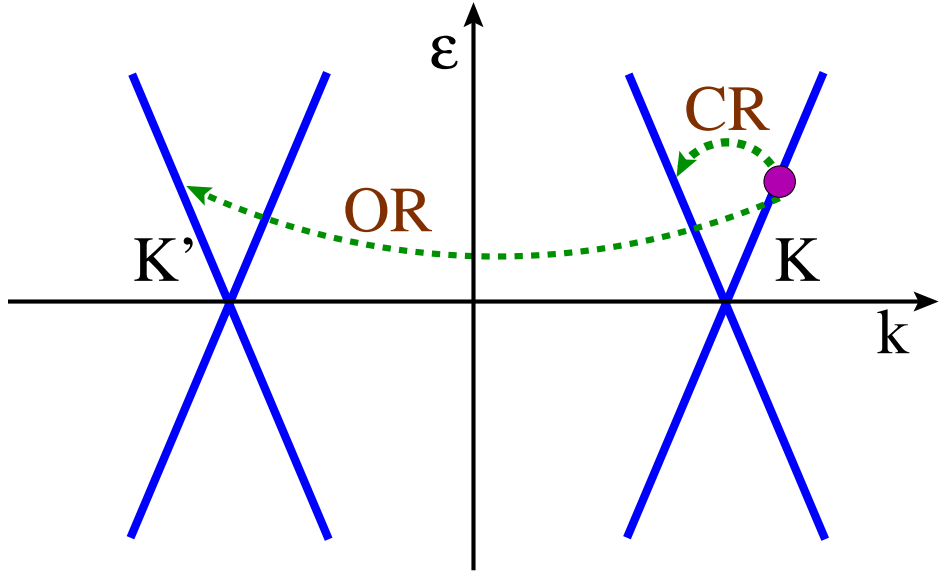} 
\caption{ {The intravalley (CR) and intervalley (OR) backscattering taking
place in the presence of the chiral barrier in the ZZ-stripe.}}
\label{Fig_2apx}
\end{figure}

In presence of the chiral barrier in the ZZ-stripe, whose height $V_{\rm B}$ is controlled by the magnitude of the local gate voltage $V_{\rm lg}$, in addition to ordinary reflection (OR) representing an inter-valley backscattering process, there appears the chiral reflection (CR) related to the intravalley backscattering as illustrated in Fig.~\ref{Fig_2apx}. The latter CR process represents an analog of Andreev reflection at the normal metal-superconductor interface.

Henceforth, we use the plane wave piece-wise approximation when the trial envelope wavefunction for $k>k_{\rm c}=1/W$ is written in the form%
\begin{eqnarray*}
\chi _{n,k}^{zz}\left( x\right) &=&t_{n}\left( 
\begin{array}{c}
\zeta _{q}^{A}\left( x\right) \\ 
\zeta _{k,q,s}^{B}\left( x\right) \\ 
0 \\ 
0%
\end{array}%
\right) e^{i\left( k+\kappa _{\varepsilon }\right) y}+r_{n}\left( 
\begin{array}{c}
\zeta _{k,q,s}^{B}\left( x\right) \\ 
\zeta _{q}^{A}\left( x\right) \\ 
0 \\ 
0%
\end{array}%
\right) e^{i\left( k-\kappa _{\varepsilon }\right) y} \\
&&+t_{n}^{\prime }\left( 
\begin{array}{c}
0 \\ 
0 \\ 
\zeta _{q}^{A}\left( x\right) \\ 
\zeta _{k,q,s}^{B}\left( x\right)%
\end{array}%
\right) e^{-i\left( k-\kappa _{\varepsilon }\right) y}+r_{n}^{\prime }\left( 
\begin{array}{c}
0 \\ 
0 \\ 
\zeta _{k,q,s}^{B}\left( x\right) \\ 
\zeta _{q}^{A}\left( x\right)%
\end{array}%
\right) e^{-i\left( k+\kappa _{\varepsilon }\right) y},
\end{eqnarray*}%
where the 1st and 2nd terms are related to the transmitted and reflected waves in the intravalley processes, while the 3rd and 4th terms represent the transmitted and reflected waves involving the ordinary intervalley scattering processes, $k$ is the longitudinal component of election momentum directed along the $\hat{y}$-axis of ZZ-stripe, $\kappa _{\varepsilon }=\sqrt{\varepsilon ^{2} - v^{2}q_{n}^{2}-\Delta ^{2}}/v$, $q_{n}$ is the transverse electron momentum component, $\Delta $ is the Stark splitting and the auxiliary functions~\cite{Fertig-1,Fertig-2} are given by 
\begin{eqnarray}
\zeta _{q}^{A}\left( x\right) &=&\sin \left( q\cdot x\right)  \nonumber \\
\zeta _{k,q,s}^{B}\left( x\right) &=&s\cdot i\left[ -q\cdot \cos \left(
q\cdot x\right) +k\cdot \sin \left( q\cdot x\right) \right] /p.
\label{Zeta}
\end{eqnarray}%
where $p=\varepsilon /v=\sqrt{k^{2}+q^{2}}$. Functions (\ref{Zeta}) describe dependence of the electron envelope wavefunction $\Phi \left( x,y\right) $ on the transversal coordinate $x$ in the ZZ-stripe sections. One can see that functions (\ref{Zeta}) conform to the edge boundary conditions for ZZ-stripe: $\zeta _{q}^{A}\left( x\right) =0$ at $x=0 $ while $\zeta _{k,q,s}^{B}\left( x\right) =0$ at $x=W$, provided Eq.~(\ref{Fert_BC_C}) holds.


\subsection{Intravalley approximation}\label{subsecA2}

We consider a simplified geometry with no conventional potential barriers separating the quantum dots from each other. Instead, in our geometry, the chiral barriers arise due to the application of the local gate potentials $V_{\mathrm{lg}} \neq 0$, or arising due to the influence of metal electrodes deposited on the top of the graphene stripe, as depicted in Fig.~\ref{Fig_1sci}. ln absence of conventional separating barriers in the quantum dots array, we neglect by the intervalley scattering such as $K\rightarrow K^{\prime }$ (or $K^{\prime }\rightarrow K$). Thus, we consider the intravalley scattering transmissions and reflection processes only. Such intravalley approximation allows a tractable analytical solution. We begin with computing reflection and transmission coefficients for the elementary $\Pi I_{G}\Pi $ block, where $I_{G}$
is the gate voltage-controlled quantum well ($V_{\mathrm{lg}}<0$) or chiral barrier ($V_{\mathrm{lg}}>0$), $W$ is the unbiased section of ZZ-stripe. The obtained coefficients are then used for finding S-matrices of more complex clusters and structures composed of many elementary $\Pi I_{G}\Pi $ blocks. In the piece-wise \emph{intravalley} approximation, the trial electron envelope wavefunction $\Psi \left( x,y\right) $ for for a single $\Pi I_{G}\Pi $\ block takes the form 
\begin{equation}
\Psi _{1}=\left( 
\begin{array}{c}
\zeta _{q_{1}}^{A}\left( x\right) \\ 
\zeta _{k_{1},q_{1},s_{1}}^{B}\left( x\right)%
\end{array}%
\right) e^{ik_{2}y}+r_{1}\left( 
\begin{array}{c}
\zeta _{-k_{1},q_{1},s_{1}}^{B}\left( x\right) \\ 
\zeta _{q_{1}}^{A}\left( x\right)%
\end{array}%
\right) e^{-ik_{1}y}  \mbox{ at } y<y_{\rm L}  \label{psi-L}
\end{equation}%
\begin{equation}
\Psi _{2}=\alpha _{2}\left( 
\begin{array}{c}
\zeta _{q_{2}}^{A}\left( x\right) \\ 
\zeta _{k_{2},q_{2},s_{2}}^{B}\left( x\right)%
\end{array}%
\right) e^{ik_{2}y}+\beta _{2}\left( 
\begin{array}{c}
\zeta _{-k_{2},q_{2},s_{2}}^{B}\left( x\right) \\ 
\zeta _{q_{2}}^{A}\left( x\right)%
\end{array}%
\right) e^{-ik_{2}y}  \mbox{ at }   y_{\rm L}<y<y_{\rm R}  \label{psi-M}
\end{equation}%
\begin{equation}
\Psi _{3}=t_{3}\left( 
\begin{array}{c}
\zeta _{q_{3}}^{A}\left( x\right) \\ 
\zeta _{k3,q_{3},s_{3}}^{B}\left( x\right)%
\end{array}%
\right) e^{ik_{3}y}   \mbox{ at }  y>y_{\rm R}  \label{psi-R}
\end{equation}%
where $r_{1}$ and $t_{3}$ are the respective reflection and transmission coefficients, indices $1$, $2$ and $3$ denote the left-, middle- (barrier or well) and right-hand neighboring sections respectively. The first term on the right-hand side in Eq.~(\ref{psi-L}) is an incident electron envelope wavefunction while the second term is the reflected wave. The first and the second terms in Eq.~(\ref{psi-M}) are the waves bouncing back and forth in the middle region. The last Eq.~(\ref{psi-R}) describes the transmitted wave. 

Technically, the coordinate dependence in the longitudinal $y$-direction is found by solving the respective boundary conditions. We assume that for the piece-wise geometry of the $\Pi I_{G}\Pi $ block, the resulting parameters are independent of the longitudinal wave vector $k$. ln the respective sections 1, 2, and 3 of the $\Pi I_{G}\Pi $ block, the trial electron envelope wavefunction $\Phi $ is represented by the plane wave solutions of Eq.~(\ref{H_Dirac}) in simplified form%
\begin{equation}
\Phi \left( x,y\right) =\left\{ 
\begin{array}{c}
\varsigma _{n,k_{1}}^{\left( +\right) }\left( x\right)
e^{ik_{1}y}+r\varsigma _{n,k_{1}}^{\left( -\right) }\left( x\right)
e^{-ik_{1}y} \mbox{ at } y<y_{\rm L} \mbox{  } (a) \\ 
\alpha \varsigma _{n,k_{2}}^{\left( +\right) }\left( x\right)
e^{ik_{2}y}+\beta \varsigma _{n,k_{2}}^{\left( -\right) }\left( x\right)
e^{-ik_{2}y} \mbox{ at }   y_{\rm L}<y<y_{\rm R} \mbox{  } (b) \\ 
t\varsigma _{n,k_{3}}^{\left( +\right) }\left( x\right) e^{ik_{3}y}  \mbox{ at }  y>y_{\rm R} \mbox{  }  (c)%
\end{array}%
\right\vert  \label{trial-f}
\end{equation}%
where the $\hat{x}$-axis is transversal (perpendicular) to the stripe axis, $0<x<W$, $W$ is the stripe width, $\hat{y}$-axis is longitudinal (along) the stripe axis, $k_{i}$ is the longitudinal electron momentum, $i=1...3$ is the index of the elementary block's section, $\alpha $, $\beta $ are the wavefunction amplitudes inside the middle section of the elementary block, $r $ is the reflection amplitude, $t$ is the transmission amplitude, and $L$ is the length of the middle section. The first term in Eq.~(\ref{trial-f}a) is the incident electron wave, while the second term is the reflected wave. The two terms in Eq.~(\ref{trial-f}b) correspond to the waves bouncing back and forth inside the middle section ($0<y<L)$, where $V=V_{0}$ while Eq.~(\ref{trial-f}c) describes the transmitted wave. In the barrier regions, the piece-wise wavevector $k_{i}$ in the i-th section is given by%
\begin{equation}
k_{i}=\sqrt{\left( \left( \varepsilon -eV_{i}\right) /\hbar v_{\mathrm{F}%
}\right) ^{2}-q_{n}^{2}},  \label{ki}
\end{equation}%
where $v_{\mathrm{F}}$ is the Fermi velocity. The geometry of the zigzag nanoribbon does not mix the valleys, and solutions near the $K$ valley with wave vector $k_{y}$ are degenerate with solutions near the $K^{\prime }$ valley with wave vector $-k_{y}$. For the $K$ valley and a given value of $k_{y} $ the nanoribbon wave functions take the form~\cite%
{Fertig-1,Fertig-2}%
\begin{equation}
\varsigma _{n,k}^{\left( s\right) }\left( x\right) =\frac{1}{C}\left( 
\begin{array}{c}
is\sinh \left( q_{n}\left( k\right) \cdot x\right) \\ 
\sinh \left[ q_{n}\left( k\right) \cdot \left( W-x\right) \right]%
\end{array}%
\right)
\end{equation}%
where the quantized transverse momentum $q_{n}\left( k\right) $ depends on $k$ (see Eq.~(\ref{Fert_BC_B})), $C$ is the appropriate normalization constant and the corresponding eigenenergies are $\varepsilon =s\gamma a_{0} \sqrt{k_{y}^{2}-q_{n}^{2}}$ with $s=\pm 1$.

An analytical solution of the boundary conditions for the single block is obtained using computer algebra. 
For the geometry of the $\Pi I_{G}\Pi $ block comprising the single barrier of length $L$ we match the wavefunction at the barrier's ends $y=y_{\mathrm{L}}$ and $y=y_{\mathrm{R}}$.
These give four linear equations for the unknown coefficients $t$, $r$, $\alpha $ and $\beta $. Then, we use Eqs.~(\ref{psi-L})-(\ref{psi-R}) to solve the boundary condition problem. Although the obtained analytical formulas are cumbersome, we use them in our numeric computation explicitly. For illustrative purposes, here we present just a simplified more compact version of the obtained formulas. For the sake of simplicity, in Eqs.~(\ref{psi-L}), (\ref{psi-M}) and (\ref{psi-R}) we set $k_{1}=k_{3}=k$, $q_{1}=q_{3}=q$, $k_{2}=k_{M}$ and $q_{2}=q_{M}$. In the above-listed assumptions, by using computer algebra, the simplified expressions for the \emph{transmission amplitude} for the chiral barrier (or well) formed in the graphene ribbon with zigzag edges take the analytical form%
\begin{equation}
t_{ZZ}=\frac{2p_{M}p q^{2}q_{M}^{2}{}e^{iL(k_{M}-k)}\csc
(qx)}{p\left( \beta _{2}\sin (qx)+q\beta _{3}\cos (qx)\right) }
\label{t-ZZ}
\end{equation}%
where 
\begin{equation}
\beta _{1}=\beta _{11}+k_{M}^{3}{}\beta _{10}- p q\cot (qx)
\label{beta-1}
\end{equation}%
\begin{eqnarray}
\beta _{11}=k_{M}^{2} p_{M} \left( e^{2i k_{M}L}-1\right)
\left( k^{2}-k q\cot (qx)+ p^2 \right) +  \nonumber \\ 
p_{M}q_{M}^{2} \left(
k^{2}e^{2i k_{M}L}-k qe^{2i k_{M} L}\cot (qx)- p^2%
\right)
\end{eqnarray}%
\begin{equation}
\beta _{10} = - ( e^{2i k_{M} L} - 1) [ (2k p - p q\cot (qx))+2k pk_{M}q_{M}^{2}]
\end{equation}%
\begin{eqnarray}
\beta _{2} &=&-2\sin ^{2}( q_{M} x)\beta _{1}+2 p_{M}q_{M}^{2}\cos ^{2}(q_{M} x)\left( kq\cot (qx)+p pe^{2i k_{M} L}-k^{2}\right)  \nonumber \\
&& + q_{M}p q\cot (qx)\left( 1-e^{2i k_{M} L}\right) \left(
k_{M}^{2}{} + q_{M}^{2} \right) \sin (2 q_{M} x)
\end{eqnarray}%
\begin{equation}
\beta _{3}=-2  p_{M}q_{M}^{2}\cos ^{2}( q_{M}x)(k-q\cot
(qx))+2\sin ( q_{M} x)\beta _{5}
\end{equation}%
\begin{eqnarray}
\beta _{4} &=&(q\cot (qx)-k)\left[ k_{M}^{2}{} p_{M} \left( e^{2i%
 k_{M} L}-1\right) + p_{M}q_{M}^{2} e^{2i k_{M} L}%
\right]  \nonumber \\
&&+\left( e^{2i k_{M} L}-1\right) \left[  k_{M}^{3}  p + k_{M}pq_{M}^{2} \right]
\end{eqnarray}%
\begin{equation}
\beta _{5}=\left( \sin ( q_{M} x)\beta _{4}- pq_{M} \left(
e^{2i k_{M} L}-1\right) \left( k_{M}^{2}+ q_{M}^{2} \right)
\cos (q_{M} x)\right) \rm{.}  \label{beta-5}
\end{equation}

In the above formulas $L$ is the length of the middle section, $p=\sqrt{k^{2}+q^{2}}$, $p_{M}=\sqrt{k_{M}^{2} + q_{M}^{2}}$ are the absolute value of electron momentum in the section $M$. The longitudinal electron momenta $k$ and $k_{M}$ are computed from the dispersion law (\ref{Fert_BC_C})%
\begin{equation}
\varepsilon _{k}^{\left( n\right) }=vp_{n}=V_{\rm G}+v\sqrt{q_{n}^{2}+k_{i}^{2}+%
\Delta _{i}^{2}/v^{2}}.
\end{equation}%

\section{Quantum dots formed on ZZ-stripes}\label{secB}

\begin{figure}
\centering
\includegraphics[width=125 mm]{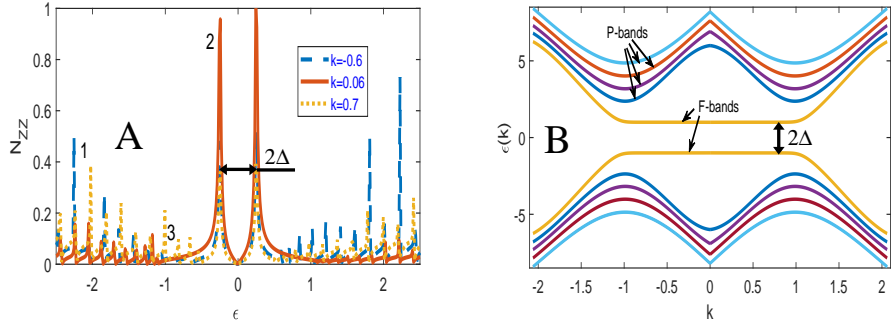} 
\caption{ {A: Electron density of states $N_{\rm ZZ}\left( \varepsilon \right) $ of ZZ-stripe for three different values of electron momentum $k$ in units of $K-K^{\prime }=4\pi /\left( 3\sqrt{3}a\right) $. The sharp peaks correspond to the quantized energy levels related to edge states. At small $k=0.06$, the lowest flat F-bands are causing sharp peaks at $\varepsilon =\pm 0.3$  (in units of Stark splitting $\Delta $). At larger $k=0.6$ and $0.7$, smaller peaks originate from the P-bands (curves 1,3). B: The electron excitation spectrum $\varepsilon \left( k\right) $ of ZZ-stripe by width $W=14$ (in units of $2/(K-K^{\prime })=3\sqrt{3}a/4\pi $) comprising energy bands, whose curvature at $k\approx 0$ is finite (P-bands) and also the two flat bands (F-bands) originating from the Stark splitting of the zero-energy level characterized by the energy gap $\Delta $,\ whose value is controlled by the split gate voltage $V_{\mathrm{sg}}$ applied as shown in Fig.~\ref{Fig_1sci}.}}
\label{Fig_3apx}
\end{figure}

In this paper we consider clusters and periodic arrays of graphene quantum dots based on graphene stripes with atomic zigzag edges (ZZ-stripes)~\cite{Swiss-ZZ,ZZ-stripe-Carbon-2019,ZZ-stripe-topolog-insul-2011,Arabs,Serhii-Graph-THz-2019,Shafr-5}. The localized state (LS) level splitting in the graphene quantum dot fabricated on the ZZ-stripe is controlled in a different way than in the device based on the stripe with armchair edges (ACh-stripe)~\cite{Sh-AQT}. To improve the quantum coherence, one exploits sharp distinctive edge state energy levels resulting from reflections at the atomic zigzag edges~\cite%
{Fertig-1,Fertig-2,Acik-Graphene-Edges-Review-2011,Swiss-ZZ,ZZ-stripe-Carbon-2019}. The levels are pronounced in the electron density of states $N_{\rm ZZ} (\varepsilon )$ as seen in Fig.~\ref{Fig_3apx}A.

Quantized energy levels related to edge states in separate sections of the quantum dot array are computed using Eq.~(\ref{Fert_BC_C}) in~\ref{secA1}. A finite electric field ${\bf E}_{\mathrm{sg}} \neq 0$ created by the split gates causes the pseudospin polarization~\cite{Shafr-Graph-Book} resulting in depletion of the electric charge on one zigzag edge and in accumulation of it on the opposite side. Then, an electric dipole is formed as soon as $eV_{\mathrm{sg}}=2\Delta /\alpha _{\mathrm{sg}}\neq 0$. When the split gate voltage $V_{\mathrm{sg}} = \vert {\bf E}_{\mathrm{sg}}\vert W$ is not applied ($V_{\mathrm{sg}} = 0$), there is a sharp singularity at zero energy. However, after applying finite $V_{\mathrm{sg}} \neq 0$, the respective zero-state energy level splits into two energy levels posing as F-bands. The level splitting mechanism is attributed to the Stark effect, as described in Sections 3.3-- 3.5 in Ref.~\cite{Shafr-Graph-Book}.  

The computed electron excitation spectrum of ZZ-stripe section is shown in Fig.~\ref{Fig_3apx}B.  The localized energy levels are obtained from the quantization condition (see calculation details in~\ref{secA1}), which is solved in respect to the excitation energy $\varepsilon _{k}^{\left(n\right) }$.  In Fig.~\ref{Fig_3apx}B one can see that at finite $V_{\mathrm{sg}} =  2\Delta /\alpha _{\mathrm{sg}} $,  in addition to conventional P-bands with finite curvature, there are two F-bands giving rise to a large electron density of states (DOS) at $\varepsilon _{k}^{\left(n\right) } = \pm \Delta$ and being associated with zigzag edge states~\cite{Shafr-Graph-Book,TEbook}. The pseudo-spin polarization driven by the $\Delta  $ term in Eq.~(\ref{Fert_BC_C}) (see~\ref{secA1}) yields an excitation spectrum $\varepsilon _{k}^{\left( n\right) }$ $=V_{\rm G}\pm \sqrt{q_{n}^{2}v^{2}+k_{i}^{2}v^{2} + \Delta _{i}^{2}}$ characterized by the energy gap $2\Delta  $ and depending on the global gate voltage $V_{\rm G}$. For a finite Stark splitting energy $\Delta  \neq 0$, ZZ-stripes are band insulators with pseudospin polarization, whose F-bands represent the highest occupied and lowest unoccupied bands. Such F-bands are characterized by a very high effective electron mass $m^{\ast }$, which, depending on the inelastic scattering rate, achieves $m^{\ast }=(10^{2}-10^{5})m_{e}>>m_{e}$. By applying the split gate voltage $V_{\rm sg}\neq 0$, one changes the value of $\Delta $ in broad region $\Delta  =\alpha _{\mathrm{sg}}eV_{\mathrm{sg} }=0.1-100$ meV, where $V_{\mathrm{sg}}$ controls the shift of the electron electrochemical potential, and actually determines whether the charge carriers inside of each respective section are electrons or holes. ln the ZZ-stripe, edge states produce much stronger spectral singularities than in graphene stripe with atomic armchair edges (ACh-stripe). As seen in Fig.~\ref{Fig_3apx}A, the respective peaks in the electron density of states $N_{ZZ}(\varepsilon )$ occurring at $\varepsilon =\pm \Delta  $ are considerably sharper (in vicinity of the gap edge they are $\propto 1/(\varepsilon -\Delta  )^{\alpha }$ where $\alpha =1.3$) than the $N_{\mathrm{ACh}}(\varepsilon )$ peaks in ACh-stripe, which are $\propto 1/\left( \varepsilon -q_{0}\right) $ and are much weaker. Numeric solutions of Eqs.~(\ref{Fert_BC}), (\ref{Fert_BC_C}) are shown in Fig.~\ref{Fig_3apx}B. One can distinguish two types of electron bands. So-called P-bands have finite curvature at $k\approx 0$ while the two flat F-bands originate from the Stark splitting of the zero-energy level characterized by the energy gap $\Delta $,\ whose value is controlled by the split gate voltage $V_{\mathrm{sg}}$ applied as shown in Fig.~\ref{Fig_1sci}.

Analytical solution (\ref{t-ZZ}), (\ref{beta-1})-(\ref{beta-5}) obtained in~\ref{secA1} describes the electron transmission through the elementary $\Pi I_{G}\Pi $ block. It is utilized to study the transport electron properties and excitation spectrum of the graphene multi-dot clusters and periodic arrays. By using Eqs.~(\ref{t-ZZ}) and (\ref{beta-1})-(\ref{beta-5}), we implement the S-matrix technique~\cite{Datta-1995} to calculate the transmission probability $T=\left\vert t\left( \varepsilon \right) \right\vert ^{2}$ and the electron spectrum $\varepsilon \left( k,q\right) $ for various geometries of quantum dot clusters and periodic arrays. 

\begin{figure}
\centering
\includegraphics[width=125 mm]{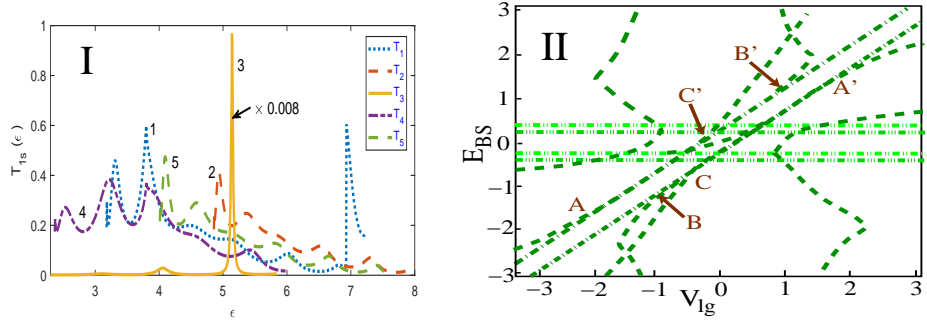} 
\caption{ {I: Energy dependence of the electron transmission probability $%
T_{1s}\left( \varepsilon \right) $ for the elementary $\Pi I_{G}\Pi $ block through a chiral barrier $I_{G}$ by height $V_{\mathrm{B}}=V_{\rm lg}=2.8$, $\delta =0.03$  (in units of $\Delta $) and $x=0.6$ in units of $1/(K-K^{\prime }) = 3\sqrt{3}a/\left(4\pi\right) $. Curves 1, 2, 4, and 5 correspond to the PP-scattering processes. In curve 3 a very sharp singularity related to the PF-scattering arises at energy $\varepsilon \approx 5.2$ (in units of $\Delta $). II: Dependence of the bound state-level energy $E_{\mathrm{BS}}$ on the local gate potential $V_{\rm lg}$, which controls the quantum well depth at $V_{\rm lg}<0$ or the barrier height at $V_{\rm lg}>0$ in the 2-dot cluster. In this way, one tunes the bound state level energies $E_{\mathrm{BS}}$ marked as A(A$^\prime $), B(B$^\prime $), C(C$^\prime $).}}
\label{Fig_4apx}
\end{figure}

In Fig.~\ref{Fig_4apx}-I, we show calculation results for the energy dependence of electron transmission probability $T_{1S}\left( \varepsilon \right) $ through the elementary $\Pi I_{G}\Pi $ block with a chiral barrier $I_{G}$ by height $V_{\mathrm{B}} = eV_{\rm lg} = 2.8$, $\delta = 0.03$ (in units of $\Delta $) at $x = 0.6$. Curves 1, 2, 4, and 5 in Fig.~\ref{Fig_3apx}-I correspond to PP-scattering processes. In curve 3 a very sharp singularity related to the PF-scattering arises at energy $\varepsilon \approx 5.2$.

An interesting issue is how the transport and spectral properties of quantum dot clusters and arrays change versus the gate voltage $V_{\rm lg}$. In particular, the $V_{\mathrm{lg}}$ magnitudes in adjacent sections control which energy levels are aligned with each other. Hence, by changing $V_{\mathrm{lg}}$, one controls the electron transitions between the energy levels in neighboring sections. Remarkably, these are confirmed by our calculation results shown in Fig.~\ref{Fig_4apx}-II. We have computed dependence of the LS energy level positions $E_{\mathrm{BS}}$ on $V_{\rm lg}$, provided the denominator of $t_{ZZ}$ in Eq.~(\ref{t-ZZ}) vanishes. One can see that by applying the electrical potential $V_{\rm lg}$ to local gates one may tune the bound state level energies $E_{\mathrm{BS}}$ and their splitting, which is illustrated in Fig.~\ref{Fig_4apx}-II where we present calculation results for a single quantum well, formed when $V_{\rm lg}<0$, which is transformed into a chiral barrier when $V_{\rm lg}>0$. In the dependence of $E_{\mathrm{BS}}$ on $V_{\rm lg}$, one can see complex behavior of the quantum dot spectrum versus the quantum well depth (or the chiral barrier height $V_{\mathrm{B}}$) and even level splitting at certain points marked as A(A'), B(B'), C(C').

\begin{figure}
\centering
\includegraphics[width=125 mm]{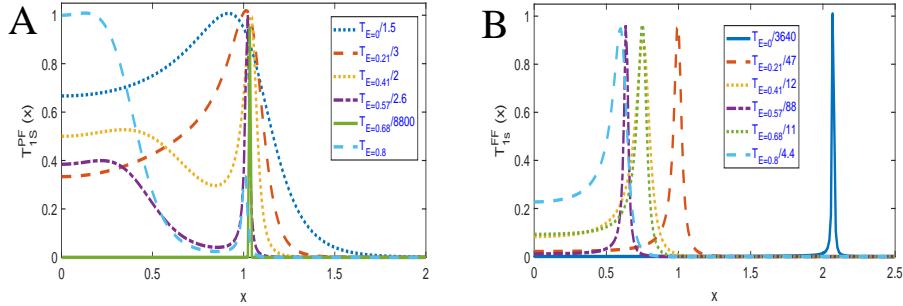} 
\caption{ {A: Spatial dependence of the transmission probability $T_{1S}^{PF}\left( x\right)$ through the single $\Pi I_{G}\Pi $ block in transversal to the ZZ-stripe axis direction for different values of electron energy $\epsilon = 0, 0.2, 0.4, 0.57, 0.68 $ and $0.8 $ in units of $\Delta $ computed for PF scattering processes. Here $V_{\rm lg} = 0.7$ and $\delta = 0.03$. Normalizing factors are shown in the inset. B:  Spatial dependence of the transmission probability $T_{1S}^{FF}\left( x\right)$  through the single $\Pi I_{G}\Pi $ block in transversal to the ZZ-stripe axis direction computed for FF scattering processes and for the same parameters as in the former Fig.~A. Normalizing factors are shown in the inset.}}
\label{Fig_5apx}
\end{figure}

The quantization in the longitudinal $y$-direction depends also on the lateral coordinate $x$, owing to the dependence of the electron wave function (\ref{Zeta}) on $x$. Solutions of the quantization condition versus $x$ for different energies are shown in Fig.~\ref{Fig_5apx}A, \ref{Fig_5apx}B, \ref{Fig_5apx}C, where we show spatial dependence of the transmission probability $T_{1S}^{PF,FF,PP}\left( x\right)$ through the single $\Pi I_{G}\Pi $ block in transversal to the ZZ-stripe axis direction for different values of electron energy $\epsilon = 0, 0.2, 0.4, 0.57, 0.68 $ and $0.8 $ in units of $\Delta $ computed for the PF, FF and PP scattering processes respectively.

\section{Coherence time in the graphene quantum dot systems}\label{secA01}
\subsection{Decoherence and dephasing of a qubit in a noisy environment}\label{secA01a}
A fundamental roadblock on the way to feasible quantum computing is the limited coherence time of qubits~\cite{ Devoret,Ithier}. The qubits couple to and thus dissipate information into the noisy environment. Longitudinal coupling describes (pure) dephasing, while the transverse coupling is responsible for relaxation. The approach~\cite{ Zanker} describes the effect of noise by mapping the noisy quantum simulator to a system of fermions coupled to a bath, similar to electron-phonon coupling. To understand the effect of decoherence and to model the transient evolution of a chain of qubits after an initialization into a non-thermal state, we consider a simple chain of qubits with dephasing due to a bosonic bath and decay due to two-level systems.
In the adiabatic limit we consider a qubit circuit subjected to pure dephasing due to a bath of harmonic oscillators characterized by its power spectral density $S (\omega ) = J_i (\omega) \coth(\beta \omega/2)$, where $ J_i (\omega)$ is the spectral function~\cite{ Zanker}. We assume that the qubits couple linearly to the displacement of the oscillators and the interaction of the quantum states with the noise is identical to a local interaction between electrons and phonons by considering the spectral density as fairly flat. For a flat spectral density the golden rule gives $\Gamma_2 \propto S(0)$.
The rates $\Gamma_i = J_i(\omega)$  correspond to the Fermi's golden-rule decay-rates for single qubits coupled to a bath of TLS and are related to the TLS spectral density $ J_i(\omega) = 2\pi \sum_s \vert g_{is} \vert^2 \delta (\omega - \omega_{is})$ ~\cite{ Zanker}. 

Given the noise spectral density $S (\omega )$ and the TLS  spectral density $ J_i(\omega )$, we evaluate the golden-rule decay rate of the qubit similarly to the case of electron-phonon interaction as described below.

\subsection{The electron state decay due to the electron-phonon interaction}\label{secA01b}
Relevant inelastic scattering mechanisms in graphene involve the electrons scattering on acoustic phonons, and on optical phonons~\cite%
{Ando-e-ph-scatter-2009,Das-Sarma-Mobility-Graphene-2008,nanomaterials-10-00039,Phonons-Graphene-Balandin-2012,TEbook,Nika,Munoz,Sanders,Savin,Karamita}. In the graphene quantum dot clusters and arrays, this process can be readily avoided by applying appropriate local gate voltage and introducing the required mismatch. Below we disregard the electron-electron collisions on the timescale of decay since the electron density in the quantum dot systems of interest is relatively low. 

For the electron-phonon scattering, the energy-dependent relaxation time $\tau \left( \varepsilon _{k}\right) $ depends on the electron density of states $N\left( \varepsilon _{k}\right) $ and temperature $T$~\cite%
{Ando-e-ph-scatter-2009,Das-Sarma-Mobility-Graphene-2008,nanomaterials-10-00039,Phonons-Graphene-Balandin-2012,TEbook,Nika,Munoz,Sanders,Savin}.
For the pristine graphene, one distinguishes several regimes~\cite{Sohier-thesis-2016} such as the Bloch-Gr\"{u}neisen (BG) regime, equipartition (EP) regime, and high temperature (HT) regime. The BG regime takes place at $0$~K$<T\leq 0.15\times T_{BG}$, $k_{B}T_{BG}=2\hbar k_{F}v_{TA/LA}$, where $v_{TA/LA}$ is the sound velocity of the TA/LA branches (typically, $v_{TA}=13.6$~km/s and $v_{LA}=21.4$~km/s). At the relevant temperatures, $k_{B}T$ is too small compared to the energy of optical phonons, thus their contribution is negligible, while the acoustic modes contribute since $k_{B}T$ is of the order of $\hbar \omega _{q,TA/LA }$. Furthermore, the occupation of initial states $f\left( \varepsilon _{k}\right) $ and scattered states $f\left( \varepsilon _{k}\pm \hbar \omega _{q,TA/LA}\right) $ are significantly different. In the EP regime at $0.15\times T_{BG}\leq T\leq $ $\hbar \omega _{A_{1}^{\prime }}/k_{B}\approx 270$ K, optical phonons do not contribute into inelastic scattering but because $\hbar \omega _{q,TA/LA} << k_{B}T <<\varepsilon _{F}$, the scattering by acoustic phonons can be approximated as elastic. In the HT regime taking place at $T\geq 0.15\times \hbar \omega _{A_{1}^{\prime }}/k_{B}\approx 270$~K, the elastic approximation for acoustic phonons is still valid. Still,  in the case of optical phonons, the three energy scales are comparable. Hence, no suitable approximation can be made globally. Since the energy of optical $A_{1}^{\prime }$ phonons is lover than the LO/TO phonons and they couple stronger, the contribution of the former is higher than the latter.

An approximate analytical expression for the relaxation rate $\gamma = \tau^{-1} _{\nu }$ is 
\begin{equation}
\gamma \left( \varepsilon _{k}\right) =\hbar \sum_{k^{\prime }}\mathcal{W}%
_{kk^{\prime }}\frac{1-f^{\left( 0\right) }\left( \mathbf{k}^{\prime
}\right) }{1-f^{\left( 0\right) }\left( \mathbf{k}\right) }\left( 1-\cos
\left( \theta _{k^{\prime }}-\theta _{k}\right) \right)  \label{tau-BG}
\end{equation}%
where $f^{\left( 0\right) }\left( \mathbf{k}\right) $ is the Fermi distribution function, $ \mathbf{k}$ is the electron  momentum. According to the Fermi rule, the electron-phonon scattering probability is%
\begin{eqnarray}
\mathcal{W}_{k^{\prime },k,\nu } &=&\frac{2\pi }{\hbar }\frac{1}{N}%
\left\vert g_{k^{\prime },k,\nu }\right\vert ^{2}\{n_{\left\vert k^{\prime
}-k\right\vert \nu } \nonumber \\
&& \times \delta \left( \varepsilon _{k^{\prime }}-\varepsilon
_{k}-\hbar \omega _{\left\vert k-k^{\prime }\right\vert ,\nu }\right) + 
\left( n_{\left\vert k-k^{\prime }\right\vert \nu }+1\right) \nonumber \\
&& \times \delta \left(
\varepsilon _{k^{\prime }}-\varepsilon _{k}+\hbar \omega _{\left\vert
k-k^{\prime }\right\vert ,\nu }\right) \}\mbox{,}  \label{P-BG}
\end{eqnarray}%
$g_{k^{\prime },k,\nu }$ is the election-phonon coupling matrix element, $n_{q\nu }$ is the Bose-Einstein distribution, $q$ is the phonon wavevector, $\nu $ is the branch index, $\omega _{q,\nu }$ is the phonon frequency and we assume that Matthiessen's rule holds and $\tau \left( \varepsilon
_{k^{\prime }}\right) \approx \tau \left( \varepsilon _{k}\right) $. In the EP and HT regimes, for the scattering on the optical $A_{1}^{\prime }$ phonons we consider only electron doping and we neglect the interband scattering, which occurs only in the case of phonon emission. Under the listed assumptions, with $\hbar \omega _{A_{1}^{\prime }}$ $=0.15$~eV, the general expression takes the form%
\begin{eqnarray}
\gamma _{A_{1}^{\prime }}^{EP,HT}\left( \varepsilon _{k}\right) &=&\frac{%
\beta _{K}^{2}}{\mu _{S}\omega _{A_{1}^{\prime }}}\{\frac{3}{2}%
n_{A_{1}^{\prime }}N\left( \varepsilon _{k}+\hbar \omega _{A_{1}^{\prime
}}\right)   \nonumber \\
&& \times \frac{1-f^{\left( 0\right) }\left( \varepsilon _{k}+\hbar \omega
_{A_{1}^{\prime }}\right) }{1-f^{\left( 0\right) }\left( \varepsilon
_{k}\right) }+\left( n_{A_{1}^{\prime }}+1\right)   \nonumber \\
&& \times N\left( \varepsilon _{k}-\hbar \omega
_{A_{1}^{\prime }}\right)    \times \frac{1-f^{\left( 0\right) }\left( \varepsilon
_{k}-\hbar \omega _{A_{1}^{\prime }}\right) }{1-f^{\left( 0\right) }\left(
\varepsilon _{k}\right) }\}\mbox{.}
\end{eqnarray}%
where $\beta _{K}=13.9$ eV/$\AA{}$ is the A$_{1}^{\prime }$ EPC parameter (GW), $\hbar \omega _{A_{1}^{\prime }}=0.15$~eV is the A$_{1}^{\prime }$ phonon energy in pristine graphene, $v_{A}=16.23$~km/s is
the effective sound velocity and $\mu_{S}=7.66$~kg/m$^{2}$ is the mass density.

Provided the optical phonons are hardly excited even at room temperature, the phonon emission process is dominant and hence the scattering probabilities in GC are given by%
\begin{equation}
\gamma \left( \varepsilon \right) =2\pi ^{2}\lambda \hbar ^{2}v^{2}N\left(
\varepsilon -\hbar \omega \right) \mbox{,}  \label{gamma}
\end{equation}%
where $N$ is normalized to $\Delta /(2\pi \hbar ^{2}v^{2})$ and%
\begin{equation}
\lambda =\frac{36\sqrt{3}}{\pi }\frac{\hbar ^{2}}{2Ma^{2}}\frac{1}{\hbar
\omega }\left( \frac{\beta }{2}\right) ^{2}  \label{lambda}
\end{equation}%
For zone-center phonons, $\omega _{\Gamma } = 196$ meV and $\lambda _{\Gamma }=2.9\times 10^{-3}(\beta _{\Gamma}/2)^{2}$, while $\omega_{K} = 161.2$~meV and $\lambda _{K}=3.5\times 10^{-3}(\beta _{K}/2)^{2}$ for zone-boundary phonons, suggesting that zone-boundary phonons dominate over zone-center phonons. Thus, the phonon frequency is the unique parameter that determines the electron lifetime~\cite{Ando-e-ph-scatter-2009}.
According to Ref.~\cite{arXiv2004-06060v1}, the average over the available data gives for pristine graphene $\lambda _{Gr}=0.22-1.1$, depending on the substrate. In pristine graphene, the electron-phonon scattering time is obtained at $T=300$ K as $\tau _{\mathrm{e-ph}}\simeq 10$ ps~\cite{Gunst}. In the graphene quantum dots, due to additional constraints on the permitted scattering processes in Eq.~(\ref{gamma}) we use the effective values $\lambda ^{\ast }=0.1$ and $2\pi \hbar ^{2}v^{2}N\left( \varepsilon -\hbar \omega \right) \simeq 10^{-6}$ eV. In the above Eqs.~(\ref{gamma})-(\ref{lambda}) we have generalized the results of Refs.~\cite{Ando-e-ph-scatter-2009,Sohier-thesis-2016} on the quantum dot geometry. Provided the optical phonons are hardly excited even at room temperature, the phonon emission process is dominant and hence the scattering rate~\cite{Ando-e-ph-scatter-2009} is estimated as $\gamma _{\mathrm{e-ph}}\simeq \allowbreak 3\times 10^{-7}$ eV, which gives $\tau _{\mathrm{e-ph}} = \hbar /\gamma \simeq 2\times 10^{-9}$ s. The obtained coherence time $\tau _{\rm c}\simeq \tau _{\mathrm{e-ph}}=2$~ns is considerably prolonged up to 10-100~ns using graphene stripes with the zigzag atomic edges.

\section{Quantum dot clusters and periodic quantum dot array}\label{subsecB1}

The transport and spectral properties of the clusters and periodic arrays of quantum dots are conveniently described in terms of the S-matrix technique~\cite{Datta-1995}. In simplest case, the "product" $\circledast $ of two S-matrices $S_{1} $ and $S_{2} $ is defined as
\begin{eqnarray}
S_{1}\circledast S_{2} &=& \left( 
\begin{array}{cc}
r_{1} & t_{1} \\ 
t_{c1} & r_{c1}%
\end{array}%
\right) \circledast \left( 
\begin{array}{cc}
r_{2} & t_{2} \\ 
t_{c2} & r_{c2}%
\end{array}%
\right)   \nonumber \\
&& = \left( 
\begin{array}{cc}
\frac{t_{c1}t_{c2}}{r_{2}r_{c1}-1} & \frac{%
t_{2}r_{c1}t_{c2}-r_{2}r_{c1}r_{c2}+r_{c2}}{r_{2}r_{c1}-1} \\ 
\frac{r_{2}t_{1}t_{c1}-r_{2}r_{1}r_{c1}+r_{1}}{r_{2}r_{c1}-1} & \frac{%
t_{1}t_{2}}{r_{2}r_{c1}-1}  \nonumber \\
\end{array}%
\right)
\end{eqnarray}%
where $r_i$ and $t_i$ are the reflection and transmission coefficients for the $i$-th scatterer respectively. The transmission probability is computed as $T_i = \vert t_i \vert^2$, while the electron excitation spectrum is obtained provided that denominator of $T_i$ vanishes.  S-matrix describing the 3GM cluster comprising three blocks is expressed as%
\begin{eqnarray}
S_{\rm 3GM} &=&\overbrace{I\Pi I_{G}\circledast \Pi I_{G}\Pi \circledast I_{G}\Pi I}\circledast 
\underbrace{\Pi I\Pi }\circledast  \nonumber \\
&&\overbrace{I\Pi I_{G}\circledast \Pi I_{G}\Pi \circledast I_{G}\Pi I}\circledast \underbrace{\Pi I\Pi }%
\circledast  \nonumber \\
&&\overbrace{I\Pi I_{G}\circledast \Pi I_{G}\Pi \circledast I_{G}\Pi I}\circledast \underbrace{\Pi I\Pi }
\label{3GM}
\end{eqnarray}%
where the interstitial $\underbrace{\Pi I\Pi }$ section is common for each of the two adjacent GM blocks. In Eq.~(\ref{3GM}), $I$ are the barriers with fixed height, and $I_{G}$ is the gate voltage-controlled barrier. One period corresponds to the GM block $I\Pi I_{G}\circledast \Pi I_{G}\Pi \circledast I_{G}\Pi I$ and
every the two adjacent GM blocks entering the GM chain are linked by the common $\underbrace{\Pi I\Pi }$ section. Results of our numeric calculations for multi-dot clusters (see Sections~\ref{sec4}) suggest that likewise to the former single dot geometry changing the chiral barrier heights in the cluster causes splitting of the quantized levels and adjusting their positions. This opens the way to a flexible all-electrical control in the graphene quantum dot systems. 

\subsection{Infinite periodic array}\label{subsecB2}
It is also interesting to see the effect of \emph{spatial coherence}, which is pronounced in an infinite periodic chain of graphene quantum dots. We apply the periodicity condition provided the electron wave function $\check{\Psi}\left( x\right) $ repeats itself in each of the GC periods $d$ of the quantum dot array by setting%
\begin{equation}
\check{\Psi}\left( x+d\right) =\lambda ^{-1}\check{S}_{P}\check{\Psi}\left(
x\right).
\end{equation}%
Hence, the periodicity condition is given in the form $\det \left( \check{1}\lambda -\check{S}_{P}\right) =0$, where $\left\vert \lambda \right\vert =1$. The S-matrix of one period $P$ in the one-dimensional geometry is%
\begin{equation}
\check{S}_{P}=\overbrace{I\Pi I_{G}\circledast \Pi I_{G}\Pi \circledast I_{G}\Pi I}\circledast 
\overbrace{\Pi I\Pi }
\end{equation}%
The last equation allows for determining the electron density of states and coherence length in 1D GC. We define the auxiliary function $\lambda =\exp \left( i\kappa d\right)$, where $\kappa_{\rm GC} $ is the Bloch quasimomentum along the $\hat{y}$-direction (i.e., along the graphene stripe axis), whose eigenvalues are obtained as%
\begin{eqnarray}
\kappa_{\rm GC} &=&\frac{1}{id}\left[ \check{S}_{P}\right] _{eigenvalues}=\frac{1}{id}%
\ln \left( 
\begin{array}{cc}
r & t \\ 
t_{c} & r_{c}%
\end{array}%
\right) _{eigenvalues}  \nonumber \\
&=&\frac{1}{2id}\ln \left( r+r_{c}\pm \sqrt{\left( r-r_{c}\right)
^{2}+4tt_{c}}\right) \simeq \frac{r+t}{id}   \label{kappa}
\end{eqnarray}%
where $t$, $t_{c}$, $r$ and $r_{c}$ are the transmission and reflections amplitudes of the one-period block. 

The electron density of states in the infinite periodic quantum dot array is computed as
\begin{equation}
N_{\mathrm{GC}}=\left\vert \frac{d\kappa _{\mathrm{GC}}}{d\varepsilon }%
\right\vert     \label{DOS}
\end{equation}%

The full S-matrix of a finite array with N periods is%
\begin{equation}
\check{S}_{NP}=\left[ \check{S}_{P}\right] ^{N}
\end{equation}%
where $N$ is the number of the $\check{S}_{P}$ blocks. One can also include the lattice "defects" $GM^{\prime }$ in the middle%
\begin{equation}
\check{S}_{NP}^{(1)}\circledast \left[ \check{S}_{GM}^{\prime }\right] ^{\ast
}\circledast \left[ \check{S}_{NP}^{(2)}\right] ^{\ast }  \label{defect}
\end{equation}%
Technically, the artificial "defects" in GC are introduced by using the
barrier $V_{\mathrm{B}}^{\prime }$ different from those $V_{\mathrm{B}}$ set in adjacent lattice sites. The lattice "defects" (\ref{defect}) alter the coherent inter-dot coupling in its vicinity which improves the flexibility of the qubit circuit design for certain applications.

Results of numeric solutions for the coherence factor $1/(d\cdot k_{\rm GC})$ in the infinite periodic quantum dot array are given in Fig.~\ref{Fig_5sci}.



\begin{thebibliography}{99}
\bibitem{Arute} F. Arute, K. Arya, R. Babbush, D. Bacon et al, Quantum supremacy using a programmable superconducting processor, Nature, {\bf 574}, 24 (2019),  https://doi.org/10.1038/s41586-019-1666-5.

\bibitem{Burnett} J. J. Burnett, A. Bengtsson, M. Scigliuzzo, D. Niepce, M. Kudra, P. Delsing and J. Bylander, Decoherence benchmarking of superconducting qubits, npj Quantum Information 5:54 (2019) ; https://doi.org/10.1038/s41534-019-0168-5.

\bibitem{Kjaergaard} M. Kjaergaard, M. E. Schwartz, J. Braumüller, P. Krantz, J. I.-J. Wang, S. Gustavsson, and W. D. Oliver, Superconducting Qubits: Current State of Play, Annual Review of Condensed Matter Physics,  {\bf 11}, 369-395 (2020).

\bibitem{Krantz} P. Krantz, M. Kjaergaard, F. Yan, T.P. Orlando, S. Gustavsson, and W. D. Oliver, A Quantum Engineer’s Guide to Superconducting Qubits, Applied Physics Reviews, {\bf  6}, 2, id.021318 (2019).

\bibitem{Zhong} H.-S. Zhong, Hui Wang, Yu-Hao Deng, Ming-Cheng Chen, Li-Chao Peng, Yi-Han Luo, Jian Qin, Dian Wu, Xing Ding, Yi Hu, Peng Hu, Xiao-Yan Yang, Wei-Jun Zhang, Hao Li, Yuxuan Li, Xiao Jiang, Lin Gan, Guangwen Yang, Lixing You, Zhen Wang, Li Li, Nai-Le Liu, Chao-Yang Lu, Jian-Wei Pan, Quantum computational advantage using photons, Science 10.1126/science.abe8770 (2020).

\bibitem{Wang} B.-C. Wang, T. Lin, H.-O. Li, S.-S. Gu, M.-B. Chen, G.-C. Guo, H.-W. Jiang, X. Hu, G. Cao, G.-P. Guo, Correlated spectrum of distant semiconductor qubits coupled by microwave photons, Science Bulletin, {\bf 66}, 4, 332-338 (2021).

\bibitem{Heuck} M. Heuck, K. Jacobs, and D. R. Englund, Controlled-Phase Gate Using Dynamically Coupled Cavities and Optical Nonlinearities, Phys. Rev. Lett. {\bf 124}, 160501 (2020).

\bibitem{Serh-Ivan-qubit-2002} S.~E.~Shafraniuk, I.~P.~Nevirkovets, J.~B.~Ketterson, A qubit device based on manipulations of Andreev bound states in double-barrier Josephson junctions, Sol. St. Comm., {\bf 121}, 9–10, 457 (2002).

\bibitem{Serhi-SISIS-qubit-2006} S.~Shafraniuk, Two-qubit gate based on a multiterminal double-barrier Josephson junction, Phys. Rev., {\bf B74}, 024521 (2006).

\bibitem{Serh-chapter-2008} S.~E.~Shafraniuk, J.~B.~Ketterson, Principles of Josephson-Junction-Based Quantum Computation, Springer-Verlag., In K.~H.~Bennemann, \& J.~B.~Ketterson (Eds.), Superconductivity: Conventional and Unconventional Superconductors, 2008, 315.

\bibitem{Geim-Chi-Tunn} M. I. Katsnelson, K. S. Novoselov \& A. K. Geim, Chiral tunnelling and the Klein paradox in graphene, Nature Physics {\bf 2}, 620--625 (2006).

\bibitem{Katsnels-Chiral-Tunnel-2012} T. Tudorovskiy, K. J. A. Reijnders and M. I. Katsnelson, Chiral tunneling in single-layer and bilayer graphene,
Phys. Scr. {\bf T146}, 014010 (2012).

\bibitem{Gold-Gordon} N. Stander, B. Huard and D. Goldhaber-Gordon, Evidence for Klein tunneling in graphene p--n junctions, Phys. Rev. Lett. {\bf 102}, 026807 (2009).

\bibitem{P-Kim} A.F.~Young and P.~Kim, Quantum interference and Klein tunnelling in graphene heterojunctions, Nature Phys., 5222 (2009).

\bibitem{Trauzettel-Spin-qubits-GQD-2007} B.~Trauzettel, D.V.~Bulaev,
D.~Loss, and G.~Burkard, Spin qubits in graphene quantum dots, Nature Phys. 
{\bf 3}, 192 (2007).

\bibitem{Shafr-Graph-Book} S. Shafranjuk, Graphene: Fundamentals, Devices,
and Applications, (Pan Stanford, 2015), SBN 9789814613477 - CAT\# N11214, 634.

\bibitem{Fertig-1} L. Brey, H.A. Fertig, Edge states and the quantized Hall
effect in graphene, Phys. Rev. B {\bf 73 (19)}, 195408-195416 (2006).

\bibitem{Fertig-2} L. Brey, H.A. Fertig, Electronic states of graphene
nanoribbons studied with the Dirac equation, Phys. Rev. B, {\bf 73 (23)},
235411-235419 (2006).

\bibitem{Acik-Graphene-Edges-Review-2011} M. Acik and Y. J. Chabal, Nature
of Graphene Edges: A Review, Jpn. J. Appl. Phys. {\bf 50} 070101 (2011).

\bibitem{Swiss-ZZ} P. Ruffieux, S. Wang, B. Yang, C. S\'{a}nchez-S\'{a}%
nchez, J. Liu, T. Dienel, L. Talirz, P. Shinde, C. A. Pignedoli, D.
Passerone, T. Dumslaff, X. Feng, K. M\"{u}llen, R. Fasel, On-surface
synthesis of graphene nanoribbons with zigzag edge topology, Nature {\bf %
531}, 489-492 (2016).

\bibitem{Sh-AQT} S.E. Shafraniuk, Graphene Quantum Dot Crystal Serving as a Multi-Qubit Circuit Operating at High Temperatures, Advanced Quantum Technologies, Corp ID: 226339422 (2020), https://doi.org/10.1002/qute.202000062

\bibitem{Datta-1995} S. Datta, Electronic Transport in Mesoscopic Systems
(Cambridge Studies in Semiconductor Physics and Microelectronic Engineering,
pp. I-Viii). Cambridge: Cambridge University Press (1995).

\bibitem{Benedek} G. Benedek, J. R. Manson, S. Miret-Artés, The electron–phonon coupling constant for single-layer graphene on metal substrates determined from He atom scattering, Phys. Chem. Chem. Phys., 23, 7575-7585 (2021).

\bibitem{PAllen} P. B. Allen, The electron-phonon coupling constant $\lambda$, in Handbook of Superconductivity, edited by C. P. Poole, Jr. (Academic Press, New York, 1999) Ch. 9, Sec. G, pp. 478-483.


\bibitem{ZZ-stripe-Carbon-2019} E. Louis, E. San-Fabian, G. Chiappe, J.A. Verges, Electron enrichment of zigzag edges of armchair--oriented graphene nano--ribbons increases their stability and induces pinning of Fermi level, Carbon, {\bf 154}, 211 (2019).

\bibitem{ZZ-stripe-topolog-insul-2011} K.-I. Imura, S. Mao, A. Yamakage and
Y. Kuramoto, Flat edge modes of graphene and of Z 2 topological insulator, Nanoscale Res. Lett. , {\bf 6}, 358 (2011).

\bibitem{Arabs} M.S.~Hossain,  F.~Al-Dirini,  F.M.~Hossain, E.~Skafidas, High Performance Graphene Nano-ribbon Thermoelectric Devices by Incorporation and Dimensional Tuning of Nanopores, Sci. Rep. , {\bf 5}, 11297 (2015).

\bibitem{Serhii-Graph-THz-2019} S.E.~Shafraniuk, Unconventional electromagnetic properties of the graphene quantum dots, Phys. Rev., {\bf B100}, 075404 (2019).

\bibitem{Shafr-5} S. E. Shafraniuk, Electromagnetic properties of the graphene junctions, European Physical Journal, B {\bf 80}, 379 (2011).

\bibitem{Zanker} S. Zanker, I. Schwenk, J.-M. Reiner, J. Lepp\"akangas, and M. Marthaler, Analyzing the spectral density of a perturbed analog quantum simulator using Keldysh formalism, Phys. Rev. B {\bf 97}, 214301 (2018).

\bibitem{Ando-e-ph-scatter-2009} H. Suzuura,T. Ando, Electron lifetime due
to optical-phonon scattering in a graphene sheet, J. Phys.: Conf. Ser.150
022080 (2009).

\bibitem{Das-Sarma-Mobility-Graphene-2008} E. H. Hwang, S. Das Sarma,
Acoustic phonon scattering limited carrier mobility in two-dimensional
extrinsic graphene, Phys. Rev. B, {\bf 77}, 11, 115449 (2008).

\bibitem{nanomaterials-10-00039} E.H. Hwang, S. Das Sarma, Screening-induced
temperature-dependent transport in two-dimensional graphene, Phys. Rev. B, 
{\bf 79}, 165404 (2009).

\bibitem{Phonons-Graphene-Balandin-2012} D. Nika, A. A. Balandin, Phonon
Transport in Graphene, Journal of Physics Condensed Matter {\bf 24} (23),
233203 (2012), DOI: 10.1088/0953-8984/24/23/233203

\bibitem{Sohier-thesis-2016} T. Sohier, Electrons and phonons in graphene:
electron-phonon coupling, screening and transport in the field effect setup.
(PhD thesis) Physics [physics]. Universit\'{e} Pierre et Marie Curie - Paris
VI, 2015.

\bibitem{arXiv2004-06060v1} G. Benedek, J. R. Manson, and S. Miret-Art\'{e}%
s, The Electron-Phonon Coupling Constant for Single-Layer Graphene on Metal
Substrates Determined from He Atom Scattering, arXiv:2004.06060v1
[cond-mat.mtrl-sci] (2020).

\bibitem{TEbook} S. Shafraniuk, Thermoelectricity and Heat Transport in
Graphene and Other 2D Nanomaterials (Elsevier - Health Sciences Division,
2016), ISBN-13: 9780323443975.

\bibitem{Nika} D.~L.~Nika, A.~A.~Balandin, Phonons and thermal transport in graphene and graphene-based materials, Rep. Prog. Phys. {\bf 80}, 036502  (2017).

\bibitem{Munoz} E. Mu\~{n}oz, J. Lu, and B. I. Yakobson, Ballistic Thermal Conductance of Graphene Ribbons, Nano Lett. {\bf 10}, 1652–1656 (2010), DOI: 10.1021/nl904206d.

\bibitem{Sanders} G.~D.~Sanders, A.~R.~T.~Nugraha, K.~Sato, J.-H. Kim, J. Kono, R.~Saito, and C.~J.~Stanton, Theory of coherent phonons in carbon nanotubes and graphene nanoribbons, J. Phys.: Condens. Matter, {\bf 25}, 144201 (2013).

\bibitem{Savin} A. V.  Savin, Y. S. Kivshar, Phononic Fano resonances in graphene nanoribbons with local defects, Scientific Reports, {\bf 7}, 4668 (2017), DOI:10.1038/s41598-017-04987-w.

\bibitem{Karamita} H. Karamitaheri, M. Pourfath, H. Kosina, and
N. Neophytou, Low-dimensional phonon transport effects in ultra-narrow, disordered graphene nanoribbons, Phys. Rev. B {\bf 91}, 165410 (2015).

\bibitem{YWang} Y. Wang, A. K. Vallabhaneni, B. Qiu, and X. Ruan, Two-dimensional thermal transport in graphene: a review of numerical modeling studies, Nanoscale and Microscale Thermophysical Engineering, {\bf 18}, 155–182, (2014).

\bibitem{Asada} M. Asada, S. Suzuki, Terahertz Emitter Using Resonant-Tunneling Diode and Applications, Sensors 2021, 21, 1384. https://doi.org/10.3390/s21041384

\bibitem{Kitagawa} S.Kitagawa, S. Suzuki, M. Asada, Wide frequency-tunable resonant tunneling diode terahertz oscillators using varactor diodes. Electron. Lett., {\bf 52}, 479–481 (2016).

\bibitem{Arzi} K. Arzi, S. Suzuki, A. Rennings, D. Erni, N. Weimann, M. Asada, W. Prost, Subharmonic injection locking for phase and frequency control of RTD-based THz oscillator. IEEE Trans. Terahertz Sci. Technol. {\bf 10}, 221–224 (2020).

\bibitem{Breuer} H.-P. Breuer, F. Petruccione, The Theory of Open Quantum Systems, (Oxford University Press, 2002). ISBN: 978-0-1985-2063-4.

\bibitem{HaugKoch} H.~Haug, S.~W.~Koch, Quantum Theory of the Optical and Electronic Properties of Semiconductors,  (World Sci. Publ. Comp., 1994), ISBN-13: 978-9810218645.

\bibitem{Nielsen} M.~A.~Nielsen, I.~L.~Chuang, Quantum Computation and Quantum Information, (Cambridge U. Press, New York, 2000), ISBN 0-521-63235-8, ISBN 0-521-63503-9

\bibitem{Wilmart2020} Q. Wilmart, M. Boukhicha, H. Graef, D. Mele, J. Palomo, M. Rosticher, T. Taniguchi, K. Watanabe, V Bouchiat, E. Baudin, J.-M. Berroir, E. Bocquillon, G. Fève, E. Pallecchi \& B. Plaçais, High-Frequency Limits of Graphene Field-Effect Transistors with Velocity Saturation, Appl. Sci., {\bf 10(2)}, 446 (2020).

\bibitem{Shaikhai} R. Shaikhaidarov, V. N. Antonov, and A. Casey, A. Kalaboukhov and S. Kubatkin, Y. Harada and K. Onomitsu, A. Tzalenchuk,  A. Sobolev, Detection of Coherent Terahertz Radiation from a High-Temperature Superconductor Josephson Junction by a Semiconductor Quantum-Dot Detector, Physical Review Applied, {\bf 5}, 024010 (2016).

\bibitem{Okamoto} T. Okamoto, N. Fujimura, L. Crespi, T. Kodera  and Y.~Kawano, Scientific Reports, Terahertz detection with an antenna-coupled highly-doped silicon quantum dot, {\bf 9}, 18574 (2019). https://doi.org/10.1038/s41598-019-54130-0.

\bibitem{Rinzan} M. Rinzan, G. Jenkins, H. D. Drew, S.~Shafraniuk, and P.~Barbara, Carbon nanotube quantum dots as highly sensitive terahertz-cooled spectrometers, Nano Lett. {\bf 12}, 6, 3097 (2012).

\bibitem{Yang} Y. Yang, G. Fedorov, S. E. Shafranjuk, T. M. Klapwijk, B. K. Cooper, R. M. Lewis, C. J. Lobb, and P. Barbara, Electronic Transport and Possible Superconductivity at Van Hove Singularities in Carbon Nanotubes, Nano Lett. {\bf 15}, 12, 7859–7866 (2015), https://doi.org/10.1021/acs.nanolett.5b02564

\bibitem{Mayle} S. Mayle, T. Gupta, S. Davis, V. Chandrasekhar, S. Shafraniuk, Thermometry and thermal management of carbon nanotube circuits, J. Appl. Phys. {\bf 117}, 194305 (2015); doi: 10.1063/1.4918667

\bibitem{Island} J. O. Island, V. Tayari, A. C. McRae, and A. R. Champagne, Few-Hundred GHz Carbon Nanotube Nanoelectromechanical Systems (NEMS), Nano Lett., {\bf 12}, 9, 4564 (2012), https://doi.org/10.1021/nl3018065.

\bibitem{Dyak} N. Dyakonova, A. El Fatimy, J. Łusakowski, W. Knap, M. I. Dyakonov, M.-A. Poisson, E. Morvan, S. Bollaert, A. Shchepetov, Y. Roelens, Ch. Gaquiere, D. Theron, and A. Cappy, Field Effect Transistors for Terahertz Detection, Appl. Phys. Lett., {\bf 88}, 141906 (2006); https://doi.org/10.1063/1.2191421.

\bibitem{Ando-2005} T.~Ando, Theory of Electronic States and Transport in Carbon Nanotubes, Journal of the Physical Society of Japan, {\bf 74}, 3, 777 (2005).

\bibitem{Devoret} M. H. Devoret and R. J. Schoelkopf, Superconducting circuits for quantum information: an outlook, Science, {\bf 339}, 1169 (2013). 

\bibitem{Ithier} G. Ithier, E. Collin, P. Joyez, P. J. Meeson, D. Vion, D. Esteve, F. Chiarello, A. Shnirman, Y. Makhlin, J. Schrie, and G. Sch\"on, Decoherence in a superconducting quantum bit circuit, Phys. Rev. B {\bf 72}, 134519 (2005).

\bibitem{Gunst} T. Gunst, T. Markussen, K. Stokbro, and M. Brandbyge, First-principles method for electron-phonon coupling and electron mobility: Applications to two-dimensional materials, Phys. Rev. B {\bf 93}, 035414 (2016).

\end{thebibliography}

\end{document}